\documentclass[amsmath,amssymb,12pt,eps,amsfonts]{article}
\usepackage{amsfonts,amssymb,amsmath,color}
\usepackage{graphicx}
\usepackage{dcolumn}
\usepackage{bm}
\newcommand{\be}{\begin{equation}}
\newcommand{\ee}{\end{equation}}
\newcommand{\beqs}{\begin{eqnarray}}
\newcommand{\eeqs}{\end{eqnarray}}

\DeclareMathOperator*{\essup}{\textrm{Essup}}
\begin{document}
\title{ \bf Existence of  Hamiltonians\\  for Some Singular Interactions on  Manifolds}

\author{\centerline {\small \c{C}a\u{g}lar Do\v{g}an$^1$, Fatih Erman$^2$, O. Teoman Turgut$^{2,\,3}$}
\\\and
 {\scriptsize{$^1$
Department of Physics, \.Istanbul University, Vezneciler, 34134,
\.Istanbul, Turkey}}
\\\and
 {\scriptsize{$^2$
Department of Physics, Bo\u{g}azi\c{c}i University, Bebek, 34342,
\.Istanbul, Turkey}}
\\\and
{\scriptsize{$^3$Feza G\"{u}rsey Institute, Kuleli Mahallesi,
\c{S}ekip Ayhan \"{O}z{\i}\c{s}{\i}k Caddesi, No: 44, Kandilli,
34684, \.Istanbul, Turkey}}
\\
{\scriptsize{Electronic mails: doganc@istanbul.edu.tr,
fatih.erman@gmail.com, turgutte@boun.edu.tr}}}

\date{\scriptsize{\textsc{\today}}}

\maketitle

%\date{\scriptsize{\textsc{\today}}}

\begin{abstract}

The existence of the Hamiltonians of the renormalized point
interactions in two and three dimensional Riemannian manifolds and
that of a relativistic extension of this model  in two dimensions
are proven. Although it is much more difficult, the proof of
existence of the Hamiltonian for the renormalized resolvent for
the non-relativistic Lee model can still be given. To accomplish
these results directly from the resolvent formula, we employ some
basic tools from the semigroup theory.
\end{abstract}

\section{Introduction}

\newtheorem{definition}{Def}
Typical field theory problems require a concept known as
renormalization, which is a way of rendering infinite quantities
to finite values to get physically meaningful results. This is a
very hard problem, and it would be interesting to find some simple
models in which the ideas can be tested in depth and a
mathematically sound description can be presented as much as
possible. This will illuminate the underlying mathematical and
physical ideas in more complicated models.

There are indeed some simple models introduced in the past. One of
them is the Dirac-delta potentials in quantum mechanics which was
first studied rigorously by Berezin and Fadeev \cite{berezin
fadeev} and later discussed extensively by Albeverio et al
\cite{Albeverio 2004, albeveriokurasov}. These works show that
Dirac-delta potential can be understood from the self-adjoint
extension point of view, hence all could be made mathematically
sound. Many body version of this problem on $\mathbb{R}^2$ is
known as the formal non-relativistic limit of the $\lambda \phi^4$
scalar field theory in (2+1) dimensions. All these are extensively
discussed first in the unpublished thesis of J. Hoppe
\cite{Hoppe}. A similar model is the non-relativistic Lee model,
which exhibits an additive divergence. We are not aware of a
mathematically rigorous discussion of this model. Physically the
relativistic simplified version of the Lee model is more important
and there is quite of a bit of work from a nonperturbative point
of view to understand the physics behind it (see the references in
\cite{kaynakturgut}). The approach we follow is introduced in
\cite{rajeev} by Rajeev, and recently we have introduced the
generalizations of these models on to manifolds \cite{altunkaynak,
nrleemodelonmanifold, teomanfatih, caglar, ermanturgutlee2}. The
rigorous understanding of the existence of the Hamiltonian left
aside in our previous works. We would like to address this issue
in the present work. There is a general approach which is exposed
in the excellent book by Albeverio and Kurasov
\cite{albeveriokurasov}, and it should be applicable in the
Dirac-delta functions case for the manifolds. However, we will use
an alternative approach. The advantage we have is the following,
the self-adjoint extension point of view becomes complicated when
we discuss field theories, it is usually hard to give a meaning to
operator valued distributions and their extension theory is even
more delicate. The other alternative which uses resolvent
convergence of cut-off Hamiltonians is problematic when we use
other regularization schemes, e.g. the powerful dimensional
regularization. This is why we want to utilize a direct approach.
In the problems that we deal with, the resulting operator is not
given but instead the resolvent is renormalized.

To answer the existence we use the following theorem taken from
semi-group theory. Let $\Delta$ be a subset of the complex plane.
A family $J(E)$, $E \in \Delta$ of bounded linear operators on the
Hilbert space $\mathcal{H}$ under consideration, which satisfies
the resolvent identity
\be J(E_1)-J(E_2)= (E_1-E_2)J(E_1)J(E_2)\; \label{resolvent
identity}\ee
for $E_1 \; ,E_2 \in \Delta$ is called a pseudo resolvent on
$\Delta$ \cite{pazy}.

The following corollary (Corollary 9.5 in \cite{pazy}) gives the
condition for which there exists a densely defined closed linear
operator $A$ such that $J(E)$ is the resolvent family of $A$: Let
$\Delta$ be an unbounded subset of $\mathbb{C}$ and $J(E)$ be a
pseudo resolvent on $\Delta$. If there is a sequence $E_k \in
\Delta$ such that $|E_k| \rightarrow \infty$ as $k\rightarrow
\infty$ and
\be \lim_{k \rightarrow \infty} - E_k J(E_k)x =x \;,
\label{resolvent limit} \ee
for all $x \in \mathcal{H}$, then $J(E)$ is the resolvent of a
unique densely defined closed operator $A$. As we will see, the
family satisfies $J(E)^{\dag} = J(E^*)$ so it is a holomorphic
family of type (A) in the sense of Kato \cite{kato}. Hence, it
defines a self-adjoint operator.

Let us mention the possibility of using the results from
\cite{albeveriokurasov} in the case of Dirac-delta functions. In
the approach of \cite{albeveriokurasov} we consider an operator
$A$ with a dense domain, and consider the singular perturbation by
an element $\phi$ in some dual space, formally:
\be A^-=A+\lambda \langle \phi  ,. \rangle \phi \;,\ee
here the bracket refers to dual pairing. The interesting case is
when  we have $\phi\in {\cal H}_{-2}(A)$, where
\be \phi\in  {\cal H}_{-2}(A) \ \ {\rm if } \ \ || {1\over 1+|A|}
\phi||< \infty \ee
and $||.||$ refers to the usual norm in the Hilbert space. Then
the theorem in \cite{albeveriokurasov} states that the operator
$A^-$ provides a self-adjoint extension with a new domain. In our
case,
\be || {1\over 1+ (-\nabla_g^2)}\delta_g(a, .)||=\int_0^\infty
\mathrm{d} s \; s \; e^{-s} \; K_s(a,a;g) < \infty \ee
thus we have the same type of singular perturbation--so called
form unbounded one. It is interesting to see how the conditions on
Ricci curvature we found will arise in this approach.

In our presentation, we do not follow a formal writing style since
the paper is rather long and has many technical details, hopefully
this makes reading more enjoyable. Various operator identities
that we use can be proved rigorously in the compact case, but they
require some more work in the noncompact case and it can be done
by using the spectral theorem for the Laplace-Beltrami operator.
We refrain from completing these arguments since they are more or
less standard in operator theory.

\section{Point Interactions in Two and Three Dimensional Riemannian Manifolds}
\label{Existence of Hamiltonian}

We adopt the natural units $\hbar=1$ in the non-relativistic
models discussed in this paper for simplicity. In
\cite{teomanfatih}, after the renormalization we have found the
resolvent kernel corresponding to the Hamiltonian for the $N$
point interactions (Dirac-delta interactions) in two and three
dimensional Riemannian manifolds as
\be R(x,y|E)= R_0(x,y|E) + \sum_{i,j=1}^{N} R_0(x,a_i|E)
\Phi^{-1}_{ij}(E) R_0(a_j,y|E) \label{resolventkernel} \;,  \ee
where
\be \label{phiheat3} \Phi_{ij} (E) =
\begin{cases}
\begin{split}
\int_0 ^\infty \mathrm{d} t \; K_{t}(a_i,a_i;g) \;
\left(e^{-t\mu_i^2} -e^{t E}\right)
\end{split}
& \textrm{if $i = j$} \\
\begin{split}
- \; \int_0^\infty \mathrm{d} t  \; K_{t}(a_i,a_j;g) \; e^{t E}
\end{split}
& \textrm{if $i \neq j$} \;.
\end{cases}
\ee
Here $\Re(E)<0$ and $K_t(x,y;g)$ is the heat kernel on the
Riemannian manifold, which is defined as the fundamental solution
to the heat equation
\be {1 \over 2m} \nabla_g^2 K_t(x,y;g) = {\partial K_t(x,y;g)
\over
\partial t} \;. \ee
In order to show that the resolvent kernel given in the equation
(\ref{resolventkernel}) corresponds to a unique densely defined
closed operator $H$, we need to first prove that it satisfies the
resolvent identity, i.e,
\be R(x,y|E_1)-R(x,y|E_2)=(E_1-E_2) \int_{\mathcal{M}}
\mathrm{d}^{D}_{g}z \; R(x,z|E_1)R(z,y|E_2)
\;.\label{resolventkernel identity } \ee
A detailed proof as well as all properties of the heat kernel that
we use in this paper, are given in our previous work
\cite{teomanfatih} and the relevant literature is also given
there. Here we will just give the main idea of the proof for the
completeness of this paper. If we substitute the equation
(\ref{resolventkernel}) into the equation (\ref{resolventkernel
identity }), we obtain
\beqs & & R_0(x,y|E_1)- R_0(x,y|E_2) + \sum_{i,j=1}^{N}
R_0(x,a_i|E_1) \Phi^{-1}_{ij}(E_1) R_0(a_j,y|E_1) \cr \hskip-2cm &
& \hspace{6cm} - \sum_{i,j=1}^{N} R_0(x,a_i|E_2)
\Phi^{-1}_{ij}(E_2) R_0(a_j,y|E_2) \cr & & = (E_1-E_2)
\int_{\mathcal{M}} \mathrm{d}^{D}_{g}z \; \Bigg[
R_0(x,z|E_1)R_0(z,y|E_2) \cr & & \hspace{3cm}+ \sum_{i,j=1}^{N}
R_0(x,z|E_1) R_0(z,a_i|E_2) \Phi^{-1}_{ij}(E_2) R_0(a_j,y|E_2) \cr
& & \hspace{3cm} + \sum_{i,j=1}^{N} R_0(x,a_i|E_1)
\Phi^{-1}_{ij}(E_1) R_0(a_j,z|E_1) R_0(z,y|E_2) \cr & & +
\sum_{i,j=1}^{N}\sum_{r,l=1}^{N} R_0(x,a_i|E_1)
\Phi^{-1}_{ij}(E_1) R_0(a_j,z|E_1) \cr & & \hspace{5cm} \times
R_0(z,a_r|E_2) \Phi^{-1}_{rl}(E_2) R_0(a_l,y|E_2) \Bigg] \;.
\label{resolvent identity prf 1} \eeqs
The term $R_0(x,y|E_1)- R_0(x,y|E_2)$ equals to the first term in
the right hand side of the equation above since the free resolvent
kernel $R_0(x,y|E)$ must satisfy the resolvent identity
(\ref{resolventkernel identity }). If we add and subtract the
terms
\be \sum_{i,j=1}^{N} R_0(x,a_i|E_1) \Phi^{-1}_{ij}(E_1)
R_0(a_j,y|E_2) \ee
\be \sum_{i,j=1}^{N} R_0(x,a_i|E_1) \Phi^{-1}_{ij}(E_2)
R_0(a_j,y|E_2) \ee
to the remaining terms in the equation above and rearrange, one
can complete the proof for the resolvent identity
(\ref{resolventkernel identity }) by showing that the difference
of the principal matrix $\Phi_{ij}(E_2)-\Phi_{ij}(E_1)$ equals to
the difference of free resolvent kernels
$R_0(a_i,a_j|E_1)-R_0(a_i,a_j|E_2)$. It is easy to show this by
using the formula expressing the free resolvent kernel as a
Laplace transformation of the heat kernel and semigroup property
of the heat kernel following a change of variable for the time
variable in the heat kernel \cite{teomanfatih}.
The equation (\ref{resolvent limit}) requires the following
condition to complete the second part of the proof
\be ||E_k R(E_k)f+f|| \rightarrow 0 \;, \ee
as $k\rightarrow \infty$, where $f$ belongs to the Hilbert space
$\mathcal{H}=L^2(\mathcal{M})$ and the norm is the usual $L^2
(\mathcal{M})$ norm. Let us choose the sequence $E_k = -k |E_0|$,
where $E_0$ is below the lower bound $E_*$ on the ground state
energy which has been found in \cite{teomanfatih}. Then, we must
show that
\be || |E_k| R(E_k)f - f || \rightarrow 0 \;, \ee
as $k \rightarrow \infty$. Using the equation
(\ref{resolventkernel}) and separating the free part, we get
\beqs  || |E_k| R(E_k)f -f || & \leq & || |E_k| R_0(E_k)f -f ||
\cr & +& |E_k| || R_0(E_k) \Phi^{-1}(E_k) R_0(E_k) f ||
\label{existence ham 2 3} \;. \eeqs
It is well known that the first part of the sum converges to zero
as $k\rightarrow \infty$, that is, the free resolvent corresponds
to a densely defined closed operator (Laplacian). Moreover, the
Laplacian on geodesically complete Riemannian manifolds is
essentially self-adjoint in $L^2(\mathcal{M})$
\cite{gaffney1,gaffney2}. Therefore, we are going to investigate
only the second term in two and three dimensions separately. Two
dimensional analysis has been already worked out in
\cite{teomanfatih} and we will just review it here and then give
the detailed proof for the three dimensional case. Since the norm
of an operator is smaller than its Hilbert-Schmidt norm: $||A||
\leq \mathrm{Tr}^{1/2}(A^{\dagger} A)$ with $A=R_0(E_k)
\Phi^{-1}(E_k) R_0(E_k)$, we have
\beqs  & &  |E_k| || R_0(E_k) \Phi^{-1}(E_k) R_0(E_k) f || \cr & &
\leq |E_k| \Bigg[ \sum_{i,j,r,l=1}^{N} \int_{\mathcal{M}}
\mathrm{d}^{D}_{g} x \; R_0(a_i,x|E_k) R_0(x,a_l|E_k) \cr & &
\hspace{4cm} \times \int_{\mathcal{M}} \mathrm{d}^{D}_{g} y \;
R_0(a_j,y|E_k) R_0(y,a_r|E_k) |\Phi_{ij}^{-1}(E_k)| \; |
\Phi_{rl}^{-1}(E_k)|\Bigg]^{1/2} \;. \label{resolvent inequality}
\eeqs
Let us first consider the diagonal case $l=i$ and $r=j$ for the
terms inside the bracket above.
\beqs  & & |E_k|\Bigg[ \sum_{i,j=1}^{N} \int_{\mathcal{M}}
\mathrm{d}^{D}_{g} x \; R_0(a_i,x|E_k) R_0(x,a_i|E_k) \cr & &
\hspace{3cm} \times \int_{\mathcal{M}} \mathrm{d}^{D}_{g} y \;
R_0(a_j,y|E_k) R_0(y,a_j|E_k)
 |\Phi_{ij}^{-1}(E_k)| \; |\Phi_{ji}^{-1}(E_k)|\Bigg]^{1/2} \cr & & \leq |E_k|\Bigg[ N^2
\underset{1\leq i \leq N} \max \; \alpha_i(E_k) \;\;
\underset{1\leq j \leq N} \max \; \alpha_j(E_k) \;\;
\underset{1\leq i,j \leq N} \max \; |\Phi_{ij}^{-1}(E_k)|^2
\Bigg]^{1/2} \;, \label{norm bound} \eeqs
where we have defined $\alpha_i(E_k)=\int_{\mathcal{M}}
\mathrm{d}^{D}_{g} y \; R_0(a_i,y|E_k) R_0(y,a_i|E_k)$ for
simplicity. It is easy to see that
\beqs  \int_{\mathcal{M}} \mathrm{d}^{D}_{g} x \; R_0(a_i,x|E_k)
R_0(x,a_l|E_k) & =& \int_{0}^{\infty}\int_{0}^{\infty} \mathrm{d}
t_1 \, \mathrm{d} t_2 \; K_{t_1 + t_2} (a_i,a_l;g) e^{- (t_1 +
t_2)|E_k|} \cr & =& \int_{0}^{\infty} \mathrm{d} t \; t \; K_{t}
(a_i,a_l;g) e^{-t |E_k|} \label{rzero2 heat kernel} \eeqs
by using the fact that the free resolvent kernel is just the
Laplace transform of the heat kernel. The upper bound of the heat
kernel was given in \cite{grigoryanbook, grigoryan} and summarized
in \cite{teomanfatih} for compact (with bounded Ricci) and
Cartan-Hadamard manifolds \cite{caglar}.
We shall use the notation for the dimensionless constants coming
from the bounds of the heat kernel as $C$ with subscripts for
simplicity since the exact form of these constants do not play any
role here. The upper bound of the heat kernel for compact (with
bounded Ricci) and Cartan-Hadamard manifolds is given in the
following form
\be \label{heatkernel bound} K_t(x,y;g) \leq
\begin{cases}
\begin{split}
\left[{C_1 \over V(\mathcal{M})} + {C_2 \over (t/2m)^{D/2}}
\right] \mbox{exp} \left(-\frac{m d^2(x,y)}{C_3 t}\right)
\end{split}
& \textrm{for compact manifolds} \\
\begin{split}
{C_4 \over (t/2m)^{D/2}} \mbox{exp} \left(-\frac{m d^2(x,y)}{C_5
t}\right)
\end{split}
& \hskip-2cm \textrm{for Cartan-Hadamard manifolds}\;,
\end{cases}
\ee
where $V(\mathcal{M})$ is the volume of the manifold and $d(x,y)$
is the geodesic distance between the point $x$ and $y$. Then,
on-diagonal upper bound of the equation (\ref{rzero2 heat kernel})
for compact manifolds (with bounded Ricci) becomes
\beqs & & \hskip-1cm \underset{1\leq i \leq N} \max \;
\alpha_i(E_k) \leq  {C_1 \over V(\mathcal{M})|E_k|^2} + C_6
(2m)^{D \over 2} \;|E_k|^{{D \over 2}-2} \;, \label{r02 compact
upper bound} \eeqs
where $C_6 = C_2 \Gamma(2-D/2)$. For Cartan-Hadamard manifolds, we
get
\beqs & & \hskip-1cm \underset{1\leq i \leq N} \max \;
\alpha_i(E_k) \leq C_4 \; (2m)^{D \over 2} \; |E_k|^{{D \over
2}-2} \;. \label{r02 CH upper bound} \eeqs
We have also
\beqs  \underset{1\leq i,j \leq N} \max \; |\Phi^{-1}_{ij}|^2 &
\leq & \underset{1\leq i \leq N} \max \; \sum_{j=1}^{N}
|\Phi^{-1}_{ij}|^2  = \underset{1\leq i \leq N} \max \;
(\Phi^{-1}(E_k)\Phi^{-1}(E_k))_{ii} \leq \rho(\Phi^{-2}(E_k)) \cr
& \leq & || \Phi^{-2}(E_k)|| \leq ||\Phi^{-1}(E_k)||^2 \eeqs
where we have used $\Phi^{\dag}(E_k)=\Phi(E_k)$ for $E_k \in
\mathbb{R}$ and $\rho$ is the spectral radius.

In order to find the upper bound for the norm of the inverse
principal matrix, we first decompose the principal matrix into two
positive matrices
\be \Phi = D - K \ee
where $D$ and $K$ stand for the on-diagonal and the off-diagonal
part of the principal matrix, respectively. Then, it is easy to
see $\Phi = D(1-D^{-1}K)$. The principal matrix is invertible if
and only if $(1-D^{-1}K)$, and $(1-D^{-1}K)$ has an inverse if the
matrix norm satisfies $||D^{-1}K ||<1$. Then, we can write the
inverse of $\Phi$ as a geometric series
\beqs \Phi^{-1}&=&(1-D^{-1}K)^{-1} D^{-1} \cr & = & \left(1 +
(D^{-1}K) + (D^{-1}K)^2+ ... \right) D^{-1} \;,\eeqs
and the norm has the following upper bound
\beqs || \Phi^{-1}|| & = & ||(1-D^{-1}K)^{-1} D^{-1}|| \leq
||(1-D^{-1}K)^{-1}|| \; || D^{-1}|| \cr & \leq & {1 \over 1-
||D^{-1}K||} \; ||D^{-1}|| \;. \eeqs
Since we are not concerned with the sharp bounds on the norm of
$\Phi^{-1}$ for this problem, we can choose $|E_k|$ sufficiently
large such that $||D^{-1}K||< 1/2$ without loss of generality and
get
\be ||\Phi^{-1}(E_k)|| \leq 2 ||D^{-1}(E_k)|| \;. \ee
Whenever $D^{-1}=
\mathrm{diag}(\Phi^{-1}_{11},\Phi^{-1}_{22},\ldots,
\Phi^{-1}_{NN})$, then
\be ||D^{-1}|| = \underset{1\leq i \leq N} \max \;
|\Phi^{-1}_{ii}| \;. \ee
The lower bound of the diagonal principal matrix for compact and
Cartan-Hadamard manifolds, which was given in \cite{teomanfatih},
leads to the upper bound of the inverse of the diagonal part of
the principal matrix. Hence, we find
\be \label{compact phi upper} \underset{1\leq i \leq N} \max \;
|\Phi_{ii}^{-1} (E_k)| \leq
\begin{cases}
\begin{split}
C_7 (2m)^{-1} \ln^{-1} \left(|E_k| /\mu^2 \right)
\end{split}
& \textrm{if $D = 2$} \\
\begin{split}
C_8 (2m)^{-3/2} \bigg[|E_k|^{1/2} - \mu \bigg]^{-1}
\end{split}
& \textrm{if $D = 3$}\;,
\end{cases}
\ee
for compact manifolds and
\be \label{cartanhadamard phi upper} \underset{1\leq i \leq N}
\max \; | \Phi_{ii}^{-1}(E_k)| \leq
\begin{cases}
\begin{split}
C_9 (2m)^{-1}\ln^{-1} \left( { |E_k| + \xi \over \mu^2 + \xi }
\right)
\end{split}
& \textrm{if $D = 2$} \\
\begin{split}
C_{10} (2m)^{-3/2} \Bigg[ \left( |E_k| + \xi \right)^{1/2} -
\left(\mu^2 +\xi \right)^{1/2} \Bigg]^{-1}
\end{split}
& \textrm{if $D = 3$} \;,
\end{cases}
\ee
for Cartan-Hadamard manifolds. Here $\xi$ is a positive constant
and defined in \cite{teomanfatih}.

If we substitute the results (\ref{r02 compact upper bound}),
(\ref{r02 CH upper bound}) and (\ref{compact phi upper}),
(\ref{cartanhadamard phi upper}) into (\ref{norm bound}) for
$D=2$, and take the limit $k \rightarrow \infty$, the result goes
to zero. Since the norm is always positive, we prove
\be |||E_k| R(E_k)f -f || \rightarrow 0 \label{resolvent norm
limit} \ee
as $k \rightarrow \infty$. For the off-diagonal terms, we do not
have to make a separate detailed analysis since one can easily
show that these terms are essentially  exponentially suppressed by
the factor $e^{-\sqrt{2m|E_k|}d(a_i,a_j)}$ due to the upper bounds
of the modified Bessel functions which are given in
\cite{teomanfatih}. Therefore, all off-diagonal terms
exponentially vanish when we take the limit $k \rightarrow
\infty$, which is enough for our purposes.

On the other hand, this proof does not work in the three
dimensional case as one can easily see. In three dimensions,
estimating the operator norm by the Hilbert-Schmidt norm  is not a
good way. Instead we will return to the second term in
(\ref{existence ham 2 3}), and show that
\beqs & & |E_k|  \Bigg[\int_{\cal M} \mathrm{d}_g^3 x \;
\sum_{i,j,r,l=1}^{N} R_0(x,a_i|E_k)\Phi^{-1}_{ij}(E_k) \int_{\cal
M} \mathrm{d}_g^3 z \;  R_0(a_j, z|E_k) f^*(z)  \cr
  &\ & \ \ \ \ \   \ \ \ \ \ \ \ \ \times  R_0(x, a_r|E_k)
  \Phi^{-1}_{rl}(E_k)\int_{\cal M} \mathrm{d}_g^3 y \; R_0(a_l,y|E_k)f(y)\Bigg]^{1/2}
\label{three ham ex} \; \eeqs
goes to zero as $k \to \infty$ for any $f\in L^2(\mathcal{M})$.
From our previous argument, we know that the inverse of the
principal matrix $\Phi$ satisfies for three dimensional compact
and Cartan-Hadamard manifolds:
\be \underset{1\leq i,j \leq N} \max \; |\Phi^{-1}_{ij} (E_k)|
\leq {C_{11}(2m)^{-3/2} \over |E_k|^{1/2}} \;, \label{inverse
principal 3D bound} \ee
where we define all the constant terms coming from the bounds of
the heat kernel as $C_{11}$ and ignore the term in the denominator
for large values of $|E_k|$ for simplicity. Moreover, we can
combine the two resolvents with the common variable $x$ in the
equation (\ref{three ham ex}). As a result, we can express this
combination as in the equation (\ref{rzero2 heat kernel}) and the
diagonal upper bounds of it for $D=3$ has been given in the
equations (\ref{r02 compact upper bound}) and (\ref{r02 CH upper
bound}) for compact and Cartan-Hadamard manifolds, respectively.
Once again, we skip the detailed calculations for the off-diagonal
terms ($i\neq r$) in the above sum since they are exponentially
suppressed by the factor $e^{-\sqrt{2m|E_k|}d(a_i,a_r)}$. We
always concentrate on the least convergent part in the terms. Once
we have achieved our goal for these terms, we are done.

Therefore, it is sufficient to deal with only the diagonal term
($r=i$) in the equation (\ref{three ham ex}). It is easy to show
that it is smaller than the following term
\beqs & & N^{3/2} |E_k| \Bigg[ \underset{1\leq i \leq N} \max \;
\alpha_i(E_k) \Bigg]^{1/2} \Bigg[ \underset{1\leq i,j \leq N} \max
\; |\Phi^{-1}_{ij} (E_k)| \underset{1\leq i,l \leq N} \max \;
|\Phi^{-1}_{il} (E_k)| \Bigg]^{1/2} \cr & & \times \Bigg[
\underset{1\leq j \leq N} \max \; \Bigg( \int_{\cal M}
\mathrm{d}_g^3 z \; R_0(a_j, z|E_k)|f(z)| \Bigg) \;
\underset{1\leq l \leq N} \max \; \Bigg( \int_{\cal M}
\mathrm{d}_g^3 y \; R_0( y, a_l|E_k)|f(y)| \Bigg) \Bigg]^{1/2} \;
. \label{r0 maxterms 3dim }\eeqs
Using the equations (\ref{r02 compact upper bound}), (\ref{r02 CH
upper bound}) and (\ref{inverse principal 3D bound}) in the above
equation, we get the upper bound of (\ref{r0 maxterms 3dim }) for
three dimensional compact manifolds
\beqs & & N^{3/2} |E_k| \Bigg[ {C_1 \over V(\mathcal{M})|E_k|^2} +
{C_6 (2m)^{3 \over 2} \over |E_k|^{1/2}} \Bigg]^{1/2} \Bigg[
{C_{11} (2m)^{-3/2}\over |E_k|^{1/2}} \Bigg] \cr & & \times \Bigg[
\underset{1\leq j \leq N} \max \; \Bigg( \int_{\cal M}
\mathrm{d}_g^3 z \; R_0(a_j, z|E_k)|f(z)| \Bigg) \;
\underset{1\leq l \leq N} \max \; \Bigg( \int_{\cal M}
\mathrm{d}_g^3 y \; R_0( y, a_l|E_k)|f(y)| \Bigg) \Bigg]^{1/2} \;
 \eeqs
and for three dimensional Cartan-Hadamard manifolds
\beqs & & N^{3/2} |E_k| \Bigg[ {C_4 (2m)^{3 \over 2} \over
|E_k|^{1/2}} \Bigg]^{1/2} \Bigg[ {C_{11} (2m)^{-3/2}\over
|E_k|^{1/2}} \Bigg] \cr & & \times \Bigg[ \underset{1\leq j \leq
N} \max \; \Bigg( \int_{\cal M} \mathrm{d}_g^3 z \; R_0(a_j,
z|E_k)|f(z)| \Bigg) \; \underset{1\leq l \leq N} \max \; \Bigg(
\int_{\cal M} \mathrm{d}_g^3 y \; R_0( y, a_l|E_k)|f(y)| \Bigg)
\Bigg]^{1/2} \; \eeqs
All these imply that the term
\be \int_{\cal M} \mathrm{d}_g^3 y \; R_0(a_j, y|E_k)|f(y)| \ee
must decay at least faster than $|E_k|^{-1/4}$. We now recall that
the free resolvent kernel is just the Laplace transform of the
heat kernel
\beqs R_0(a_j, y|E_k)= \; \int_0^\infty \mathrm{d} t \; e^{-t
|E_k|} K_t(a_j,y;g)  \;, \eeqs
so that we can find an upper bound for it by using the equation
(\ref{heatkernel bound}) and evaluating the integrals over $t$
\beqs & & R_0(a_j, y|E_k) \leq {m C_{12} \over d(a_j,y)} \exp
\Bigg[ -2 \Bigg( {m d^2(a_j,y)|E_k| \over C_3} \Bigg)^{1/2} \Bigg]
\cr & & \hspace{1cm} + {C_{13} d(a_j,y) \sqrt{m} \over
V(\mathcal{M}) \sqrt{|E_k|}} \Bigg[1+ \Bigg({C_3 \over m
d^2(a_j,y)|E_k|} \Bigg)^{1/2} \Bigg] \exp \Bigg[ -2 \Bigg( {m
d^2(a_j,y)|E_k| \over C_3} \Bigg)^{1/2} \Bigg] \;, \eeqs
for three dimensional compact manifolds and
\beqs R_0(a_j, y|E_k) \leq {m C_{14} \over d(a_j,y)} \exp \Bigg[
-2 \Bigg( {m d^2(a_j,y)|E_k| \over C_5} \Bigg)^{1/2} \Bigg] \;,
\eeqs
for three dimensional Cartan-Hadamard manifolds. Here we have used
the upper bound of the modified Bessel function $K_1(x)$ given in
\cite{teomanfatih}.

For simplicity, let us first consider the generic term which is
common for both compact and Cartan-Hadamard manifolds and keep
the inverse volume term aside for the moment. Then, we find for
the generic term
\be \int_{\cal M}  \mathrm{d}_g^3 y \; R_0(a_j, y|E_k)|f(y)| \leq
m C_{12} \int_{\cal M} \mathrm{d}_g^3 y \; \exp \bigg[{-2 \left(
{m d^2(a_j,y) |E_k| \over C_3}\right)^{1/2}} \bigg] {|f(y)|\over
d(a_j,y)} \;.  \ee
We now divide the integration region into two pieces
\beqs & & \int_{B_\delta(a_j)} \mathrm{d}_g^3 y \; \exp \Big[{-2
\left( {m d^2(a_j,y) |E_k| \over C_3}\right)^{1/2}} \Big]
{|f(y)|\over d(a_j,y)} \cr & & \quad + \int_{{\cal{M}}\setminus
B_\delta(a_j)} \mathrm{d}_g^3 y \; \exp \Big[{-2 \left( {m
d^2(a_j,y) |E_k| \over C_3}\right)^{1/2}} \Big] {|f(y)|\over
d(a_j,y)} \label{two terms} \;, \eeqs
where $B_{\delta}(a_j)$ is the geodesic ball of radius $\delta$
centered at $a_j$. It is easily seen that
\beqs & & \int_{{\cal{M}}\setminus B_\delta(a_j)} \mathrm{d}_g^3 y
\; \exp \Big[{-2 \left( {m d^2(a_j,y) |E_k| \over
C_3}\right)^{1/2}} \Big] {|f(y)|\over d(a_j,y)} \cr & & \leq {1
\over \delta} \exp \Big[{- \left( {m \delta^2 |E_k| \over
C_3}\right)^{1/2}} \Big] \int_{{\cal{M}}\setminus B_\delta(a_j)}
\mathrm{d}_g^3 y \; \exp \Big[{-\left( {m d^2(a_j,y) |E_k| \over
C_3}\right)^{1/2}} \Big] |f(y)|\cr & & \leq {1 \over \delta} \exp
\Big[{- \left( {m \delta^2 |E_k| \over C_3}\right)^{1/2}} \Big]
\int_{{\cal{M}}} \mathrm{d}_g^3 y \; \exp \Big[{-\left( {m
d^2(a_j,y) |E_k| \over C_3}\right)^{1/2}} \Big] |f(y)| \cr & &
\leq {1 \over \delta} \exp \Big[{- \left( {m \delta^2 |E_k| \over
C_3}\right)^{1/2}} \Big] \Bigg[\int_{\cal{M}} \mathrm{d}_g^3 y \;
\exp \Big[{-2 \left( {m d^2(a_j,y) |E_k| \over C_3}\right)^{1/2}}
\Big] \Bigg]^{1/2} ||f|| \;, \label{2nd term} \eeqs
where we have used the fact that $d(a_j,y) \geq \delta$ for all
$j$ and $y\in {\cal{M}}\setminus B_\delta(a_j)$ in the second
line. We then find an upper bound in terms of the norm of the
function $f(x)$ by using Cauchy-Schwartz inequality in the last
line.

For compact manifolds, it is a simple matter to find the upper
bound to the above integral
\beqs & & \int_{{\cal{M}}\setminus B_\delta(a_j)} \mathrm{d}_g^3 y
\; \exp \Big[{-2 \left( {m d^2(a_j,y) |E_k| \over
C_3}\right)^{1/2}} \Big] {|f(y)|\over d(a_j,y)} \cr & & \leq {1
\over \delta} \exp \Big[{- \left( {m \delta^2 |E_k| \over
C_2}\right)^{1/2}} \Big] \Bigg[V(\mathcal{M}) \; \underset{y\in
\mathcal{M}} \sup \; \Bigg( \exp \Big[{-2 \left( {m d^2(a_j,y)
|E_k| \over C_3}\right)^{1/2}} \Big] \Bigg) \Bigg]^{1/2}||f|| \cr
& & \leq {1 \over \delta} \exp \Big[{- \left( {m \delta^2 |E_k|
\over C_2}\right)^{1/2}} \Big] V^{1/2}(\mathcal{M})||f|| \;, \eeqs
due to the fact that the volume of a compact manifold is finite.
For non-compact manifolds, it is useful to consider the above
integral in the Riemann normal coordinates near one of the centers
$a_i$. We further assume that the radius of the ball $\delta$ is
less than the injectivity radius $\mathrm{inj}(a_i)$. Let us
recall that in Gaussian spherical coordinates, the volume integral
of a function $h$ on a $D$ dimensional Riemannian manifold
$\mathcal{M}$ can be written as
\be \int_{\cal M} \mathrm{d}_g^D x \; h(x) =
\int_{\mathbb{S}^{D-1}} \mathrm{d}\Omega \int_0^{\rho_\Omega}
\mathrm{d}r \; r^{D-1} J(r,\theta) h(r,\theta) \; . \ee
Here $\theta = (\theta_1,\ldots \theta_{D-1})$ denotes the
direction in the tangent space around a point that we choose, and
$\rho_\Omega$ refers to distance to the cut locus of the point in
the direction $\theta$. Hence, we get:
\beqs & & \int_{\cal{M}} \mathrm{d}_g^3 y \; \exp \Big[{-2 \left(
{m d^2(a_j,y) |E_k| \over C_3}\right)^{1/2}} \cr & & \hspace{3cm}=
 \int_{\mathbb{S}^2} \mathrm{d}\Omega \int_0^{\rho_\Omega} \mathrm{d}r \;
 r^2 J(r,\theta) \exp \Big[ {-2 \left({m r^2
|E_k| \over C_3}\right)^{1/2}}\Big]  \;. \label{gaussian spherical
coord} \eeqs
To proceed further we assume that ${\cal M}$ has Ricci tensor
bounded from below by $K_1$, i.e. $ {\rm Ric}(.,.)> K_1 \; g(.,.)$
everywhere and the sectional curvature is bounded from above by
$K_2$ on $\overline{B_{\delta}(a_i)}$. The upper bound on the
sectional curvature is automatically satisfied, since there are a
finite number of Dirac-delta centers and because we take the
metric to be $\mathcal{C}^\infty(\mathcal{M})$. Had we considered
a random distribution of Dirac-delta function centers, in which
case they could have been located at arbitrarily distant points
where the sectional curvature could have been unbounded, we would
have had to constrain the sectional curvature from above. Then
Bishop-Gunther volume comparison theorems state that the Jacobian
factor of the Gaussian spherical coordinates in $D$ dimensions
satisfies \cite{gallot,chavel2},
\be {\textrm{sn}^{D-1}_{K_2}(r)\over r^{D-1}} \leq J(r,\theta)
\leq {\textrm{sn}^{D-1}_{K_1}(r)\over r^{D-1}} \; , \label{J} \ee
where
\be \textrm{sn}_K(r)=
\begin{cases}
{\sin(\sqrt{K} r) \over \sqrt{K}}& {\rm if} \ K > 0\\
r & {\rm if} \ K=0\\
{\sinh(\sqrt{-K} r) \over \sqrt{-K}}& {\rm if} \ K < 0 \;. \\
 \end{cases} \label{sn} \ee
Then the upper bound of the equation (\ref{gaussian spherical
coord}) becomes
\beqs & & \int_{\mathbb{S}^2} \mathrm{d}\Omega \int_0^\infty
\mathrm{d} r  {\sinh^2(\sqrt{-K_1}r) \over (-K_1)} \exp \Big[ {-2
\left({m r^2 |E_k| \over C_3}\right)^{1/2}}\Big] \cr & & =
\frac{1}{|K_1|^{3/2}} \int_{\mathbb{S}^2} \mathrm{d}\Omega
\int_0^\infty \mathrm{d} r^\prime  \; \sinh^2(r^\prime) \exp \Big[
{-2 \left({m {r^\prime}^2 |E_k| \over C_3
|K_1|}\right)^{1/2}}\Big] \cr & & \leq {\pi \over |K_1|^{3/2}}
\int_0^\infty \mathrm{d} r^\prime \; \exp \Bigg[ {-2 r^\prime
\bigg( \left({m |E_k| \over C_3 |K_1|}\right)^{1/2}} - 1 \bigg)
\Bigg] \cr & & = {\pi \over 2|K_1|^{3/2} \bigg[ \left({m |E_k|
\over C_3 |K_1|}\right)^{1/2} - 1 \bigg]} \;. \eeqs
as long as $\left({m |E_k| \over C_3 |K_1|}\right)^{1/2} \geq 1$.
Since we are interested in the limit $k \rightarrow \infty$ it is
satisfied for sufficiently large values of $|E_k|$. Therefore,
equation (\ref{2nd term}) is smaller than
\beqs &\ & {\Big({\pi \over 2}\Big)^{1/2}  \over \Big( \delta
|K_1|^{3/4} \Big)} \exp \Bigg[{- \left( {m \delta^2 |E_k| \over
C_3}\right)^{1/2}} \Bigg] \bigg[ \left({m |E_k| \over C_3
|K_1|}\right)^{1/2} - 1 \bigg]^{-1/2} ||f|| \;. \label{2nd
termresult} \eeqs
If we choose $\delta=(m\sqrt{R})^{-1/3}|E_k|^{-1/3}$, where $R$ is
an appropriate scale coming from the Ricci tensor around a point,
where Ricci tensor is non-zero. The prefactor multiplying the
exponent goes to infinity whereas the exponent decays rapidly. In
fact, it decays fast enough to make the expression as a whole go
to zero as $k \to \infty$.

Let us go back to the first integral in the equation (\ref{two
terms}) and write it in the Gaussian spherical coordinates:
\be \begin{split} \int_{\mathbb{S}^2} \mathrm{d}\Omega
\int_0^\delta \mathrm{d} r & \; r^2 J(r,\theta) \exp \Big[ {-2
\left({m r^2 |E_k| \over C_3}\right)^{1/2}}\Big]
{|f(r,\theta)|\over r} \; .
\end{split}\label{deltaDelta}  \ee
Let us now make the observation that there are constants
$A_+,A_-$, which depend only on $\delta$ and $K_i$'s such that,
\be
 A_-(K_i,K_j) \leq {{\rm sn}_{K_i}(r)\over {\rm
 sn}_{K_j}(r)} \leq A_+(K_i,K_j)\label{Aplus Aminus}
\ee
for $r\in [0,\delta]$. For this part of the integral we use the
following characterization of essential supremum: let us define
\be
 \Lambda(\epsilon)=\mu(\{r\in [0, \delta]| \ |r^{3/2}F(r)|>\epsilon\})
 ,\ee
where $\mu$ is the standard Lebesgue measure. Then we have
\be
 \essup_{r \in [0, \delta]}|r^{3/2}F(r)|=
 \inf_\epsilon \{\epsilon| \Lambda(\epsilon)=0\}
\;. \ee
Let us use now $F(r)=\int_{\mathbb{S}^2} \mathrm{d}\Omega
|f(r,\theta)|$, and using Bishop-Gunther bound  for the first part
as,
\be \int_0^\delta \mathrm{d} r \; r^{3/2} \int_{\mathbb{S}^2}
\mathrm{d}\Omega \; |f(r,\theta)| { \exp \Big[{-2 \left({m r^2
|E_k| \over C_3}\right)^{1/2}} \Big] \over r^{1/2}}
{\textrm{sn}^2_{K_1}(r)\over r^2} \ee
which is smaller than;
\beqs &\ & A_+^2(K_1,0)\int_0^\delta \mathrm{d} r \; r^{3/2}
\int_{\mathbb{S}^2} \mathrm{d}\Omega \; |f(r,\theta)| { \exp
\Big[{-2 \left({m r^2 |E_k| \over  C_3}\right)^{1/2}} \Big] \over
r^{1/2}}\cr &\ & \ \ \ \ \ \leq A_+^2(K_1,0)\Big(\essup_{r\in
[0,\delta]}|r^{3/2}F(r)|\Big)\Bigg( \int_0^\delta \mathrm{d} r \;
{\exp \Big[{-2 \left({m r^2 |E_k| \over C_3}\right)^{1/2}} \Big]
\over r^{1/2}} \Bigg)\cr  &\ & \ \ \ \ \
 \leq \Big(\essup_{r\in [0,\delta]}|r^{3/2}F(r)|\Big) {A_+^2(K_1,0) \over
2(m/ C_3)^{1/4} |E_k|^{1/4}} \eeqs
If we take the limit $\delta=(m\sqrt{R})^{-1/3}|E_k|^{-1/3}\to 0$,
we claim that the essential-suppremum goes to zero. To see this,
we observe by Markov inequality \cite{stein} that
\beqs \Lambda(\epsilon)& \leq &{1\over \epsilon} \int_0^\delta
\mathrm{d} r \; |r^{3/2}F(r)| \cr & \leq & {1\over \epsilon}\Bigg[
\int_0^\delta \mathrm{d}r \; r \Bigg]^{1/2} \Bigg[\int_0^\delta
\mathrm{d}r \; r^2 \left(\int_{\mathbb{S}^2} \mathrm{d}\Omega \;
|f(r,\theta)|\right)^2 \Bigg]^{1/2}\cr & \leq & {1\over
\epsilon}{\delta\over \sqrt{2}} \Bigg[\int_0^\delta \mathrm{d} r
\; {r^2\over \textrm{sn}^2_{K_2}(r)}
\textrm{sn}^2_{K_2}(r)\int_{\mathbb{S}^2} \mathrm{d} \Omega \;
|f(r,\theta)|^2 \int_{\mathbb{S}^2}\mathrm{d} \Omega \Bigg]^{1/2}
\cr & \leq & {1\over \epsilon}{\delta\over \sqrt{2}}(4\pi)^{1/2}
A_+(0, K_2)\Bigg[\int_0^\delta \mathrm{d} r \;
\textrm{sn}^2_{K_2}(r)\int_{\mathbb{S}^2} \mathrm{d} \Omega \;
|f(r,\theta)|^2\Bigg]^{1/2} \cr & \leq & {1\over
\epsilon}{\delta\over \sqrt{2}}(4\pi)^{1/2} A_+(0,
K_2)\Bigg[\int_0^\delta \mathrm{d} r \; r^2 \int_{\mathbb{S}^2}
\mathrm{d}\Omega \; J(r,\theta) |f(r,\theta)|^2\Bigg]^{1/2} \cr
 &\leq &  {1\over \epsilon}{\delta\over \sqrt{2}}(4\pi)^{1/2} A_+(0, K_2)||f|| \;. \eeqs
For any $\epsilon>0$, our choice of $\delta$ implies that we can
make $\Lambda(\epsilon)=0$. Thus, the infimum goes to zero in the
limit as $k \rightarrow \infty$. As a result we see that the
equation (\ref{three ham ex}) is
 smaller than
 \be \begin{split}
 & C_{15} \Bigg[ {A_+^2(K_1,0) \over 2(1/C_3)^{1/4}} \Big(\essup_{r\in [0,\delta]}|r^{3/2}F(r)|\Big) \\ &
  +|E_k|^{1/4} m^{1/4}
{\exp\Big[ {-\left({m \delta^2 |E_k| \over C_3} \right)^{1/2}
}\Big] \over \Big(\delta |K_1|^{3/4}\Big)} \Big({\pi \over
2}\Big)^{1/2} \bigg[ \left({m |E_k| \over C_3 |K_1|}\right)^{1/2}
- 1 \bigg]^{-1/2}||f|| \Bigg]
  \;, \end{split} \ee
and it goes to zero as $k \to \infty $. The repeated application
of the same analysis leads us to the same conclusion for the other
terms coming from the inverse volume term which has been omitted
for simplicity. Indeed, all these terms decay with $|E_k|$ faster
than the result that we have obtained above. This completes the
proof of the existence of the Hamiltonian for point interactions
in three dimensional Riemannian manifolds.

\section{Relativistic Point Interactions on Two Dimensional Riemannian Manifolds}

The resolvent for this system have been found in \cite{caglar} and
it is given by
\be \label{greeneq} R(E)=R_0(E)+R_0(E) b^\dagger \Phi^{-1}(E) b
R_0(E) \;, \ee
where
\be \label{offdiag} b^\dagger = \sum_{i=1}^N \phi^{(-)}(a_i)
\chi_i  \ee
and
\begin{eqnarray}
\label{regularphi} \nonumber \Phi(E) &=& \frac{1}{\sqrt{\pi}}
\sum_{i=1}^N \int_0^{\infty} \mathrm{d} s \; e^{-s^2/4}
\int_0^{\infty} \mathrm{d} u \; \left( e^{s \mu_i \sqrt{u}}-e^{s E
\sqrt{u}} \right) e^{-u m^2}
K_u(a_i,a_i;g) \chi^{\dag}_{i} \chi_i \\
&-& \frac{1}{\sqrt{\pi}} \sum_{i,j \atop (i\neq j)}
\int_0^{\infty} \mathrm{d} s \; e^{-s^2/4} \int_0^{\infty}
\mathrm{d} u \; e^{sE\sqrt{u}} e^{-um^2} K_u(a_i,a_j;g)
\chi^{\dag}_{i} \chi_j
\end{eqnarray}
Here $\phi^{(-)}(x)$ is the positive frequency part of the bosonic
field and $a_i$ stands for the position of one of the $N$
Dirac-delta function potential centers and $\mu_i$ is the bound
state energy for the single delta center at $a_i$. The operator
$\chi_i$, called angel operator, was first introduced for this
purpose by Rajeev in \cite{rajeev} and it obeys orthofermionic
algebra. $K_t(x,y;g)$ is the heat kernel on the Riemannian
manifold, which is defined as the fundamental solution to the heat
equation

\be \nabla_g^2 K_t(x,y;g) = {\partial K_t(x,y;g) \over
\partial t} \;. \ee

We would like to first check whether $R(E)$ satisfies the
resolvent identity (\ref{resolvent identity}).
If we put the form of the resolvent in second quantized form,
\be R(E)=R_0(E)+R_0(E) b^\dagger \Phi^{-1}(E) b R_0(E) \ee
into the above resolvent identity (\ref{resolvent identity}) and
we simplify by purely algebraic operations, to arrive at the
following identity,
\be
\Phi_{ij}(E_1)-\Phi_{ij}(E_2)+b_i(R_0(E_1)-R_0(E_2))b_j^\dagger=0
\;, \label{required result} \ee
where we stripped off the angels and wrote everything in terms of
explicit matrix indices, and thus
$\Phi(E)=\Phi_{ij}(E)\chi^\dag_i\chi_j$ and also
$b_i=\phi^{(+)}(a_i)$ and similarly for $b_j^\dag$. Let us now
verify the above identity, we note that
\be \Phi_{ij}(E_1)-\Phi_{ij}(E_2) =
 \frac{1}{\sqrt{\pi}}
\int_0^{\infty} \mathrm{d} s \; e^{-s^2/4} \int_0^{\infty}
\mathrm{d} u \; [e^{sE_2\sqrt{u}}-e^{sE_1\sqrt{u}}] e^{-um^2}
K_u(a_i,a_j;g) \;. \ee
Let us work out the other term, acting on {\it no particle Fock
space}, this is the same calculation we have done for the
renormalized term. For simplicity, we present the calculation in a
formal eigenfunction expansion of the Laplace operator (which is
rigorously valid for only compact manifolds)
\beqs \nonumber & & b_i(R_0(E_1)-R_0(E_2))b_j^\dagger =
\sum_\sigma f^*_\sigma(a_i){a_\sigma\over
\sqrt{\omega_\sigma}}\Big[{1\over H_0-E_1}-{1\over H_0-E_2}\Big]
\sum_\lambda
f_\lambda(a_j){a_\lambda^\dagger\over \sqrt{\omega_\lambda}}\\
\nonumber &=& \sum_\sigma f^*_\sigma(a_i){a_\sigma\over
\sqrt{\omega_\sigma}}\sum_\lambda
f_\lambda(a_j){a_\lambda^\dagger\over
\sqrt{\omega_\lambda}}\Big[{1\over H_0+\omega_\lambda-E_1}-{1\over
H_0+\omega_\lambda-E_2}\Big]
\\ \nonumber
&=& \sum_\sigma \sum_\lambda
f^*_\sigma(a_i)f_\lambda(a_j)\Big[{a_\lambda^\dagger\over
\sqrt{\omega_\lambda}} {a_\sigma\over \sqrt{\omega_\sigma}}+
{\delta_{\sigma\lambda} \over \sqrt{\omega_\lambda}
\sqrt{\omega_\sigma}}\Big] \Big[{1\over
\omega_\lambda-E_1}-{1\over \omega_\lambda-E_2}\Big]
\\ \nonumber
&=& \sum_\lambda f^*_\lambda(a_i)f_\lambda(a_j){1 \over
\omega_\lambda} \Big[{1\over \omega_\lambda-E_1}-{1\over
\omega_\lambda-E_2} \Big]\\ \nonumber &=&\int_0^\infty \mathrm{d}
s \; s \sum_\lambda f^*_\lambda(a_i)f_\lambda(a_j)\int_0^1
\mathrm{d}
\zeta \; e^{-s\omega_\lambda} \Big[ e^{s\zeta E_1}-e^{s\zeta E_2}\Big] \\
\nonumber &=&\int_0^\infty \mathrm{d} s \; s \int_0^1 \mathrm{d}
\zeta \; {s\over 2 \sqrt{\pi}}\int_0^\infty {\mathrm{d} u\over
u^{3/2}}
 e^{-s^2/4u-m^2 u} K_u(a_i,a_j;g)
 \Big[ e^{s\zeta E_1}-e^{s\zeta E_2}\Big] \\
 \nonumber
&=& \int_0^\infty  \mathrm{d} s \;s \int_0^1 \mathrm{d} \zeta \;
{s\over 2 \sqrt{\pi}} \int_0^\infty \mathrm{d} u \; e^{-s^2/4-m^2
u} K_u(a_i,a_j;g)\Big[
e^{s\sqrt{u}\zeta E_1}-e^{s\sqrt{u} \zeta E_2}\Big] \\
&=&\int_0^\infty \mathrm{d} s \; {1\over \sqrt{\pi}}\int_0^\infty
\mathrm{d} u \; e^{-s^2/4-m^2 u} K_u(a_i,a_j;g)\Big[ e^{s\sqrt{u}
E_1}-e^{s\sqrt{u} E_2}\Big] \;. \eeqs
After calculating the $\zeta$ integral, we performed an
integration by parts over the variable $s$. Hence we have found
the required result (\ref{required result}).

Similar to the previous problem, let us choose the sequence $E_k =
-k |E_0 | = -|E_k|$, where $E_0$ is sufficiently below the lower
bound $E_*$ on the ground state energy which has been found in
\cite{caglar} and negative. We now want to show (\ref{resolvent
norm limit}). Substituting the resolvent equation, written in the
second quantized language, in this expression we obtain the
following:
\be \lim_{k \rightarrow \infty}||E_k
[(H_0-E_k)^{-1}+(H_0-E_k)^{-1}\phi^{(-)}(a_i) \Phi^{-1}_{ij}(E_k)
\phi^{(+)}(a_j)(H_0-E_k)^{-1}]f+f||=0 \;. \ee
The free resolvent, that is the first term in the above equation,
\be \lim_{k \rightarrow \infty}||E_k R_0(E_k)f+f||=\lim_{k
\rightarrow \infty}||E_k (H_0-E_k)^{-1}f+f||=0 \ee
already satisfies the resolvent equation hence we should look only
into the second part. To see this, note that by the triangle
inequality
\beqs & & \lim_{k \rightarrow \infty} ||E_k R(E_k)f+f|| \leq
\lim_{k \rightarrow \infty} \left( ||E_k (H_0-E_k)^{-1}f+f||
\right. \cr & & + \left. ||E_k [(H_0-E_k)^{-1}\phi^{(-)}(a_i)
\Phi^{-1}_{ij}(E_k) \phi^{(+)}(a_j)(H_0-E_k)^{-1}]f|| \right) \cr
& & \leq \lim_{k \rightarrow \infty} ||E_k
[(H_0-E_k)^{-1}\phi^{(-)}(a_i) \Phi^{-1}_{ij}(E_k)
\phi^{(+)}(a_j)(H_0-E_k)^{-1}]f|| \;. \eeqs
We choose a one-particle wave function of the form given below.
Even though, this is the most general one-particle wave function
one can write down, it does not include multi-particle wave
functions. However, due to the mutually non-interacting nature of
the particles involved, the total Hamiltonian appearing in the
resolvent will be a sum of $n$ identical, individual Hamiltonians
in the case of a $n$-particle state and therefore will decay
faster than in the one-particle case.
\be |\psi \rangle = \int_{\mathcal{M}}  \mathrm{d}_{g}^{2} x
\;\psi(x) \phi^{(-)}(x)|0 \rangle = \sum_\sigma \hat
\psi(\sigma){a^\dagger_\sigma \over \sqrt \omega_\sigma}|0 \rangle
\;.  \ee
A direct computation now reveals that,
\be \langle \psi|\psi \rangle =\sum_\sigma {|\hat
\psi(\sigma)|^2\over \omega_\sigma} \;. \ee
We verify that the limit
\beqs \lim_{k \rightarrow \infty}|E_k| \,
||\sum_{i=1}^{N}\sum_{j=1}^{N} (H_0+|E_k|)^{-1}\phi^{(-)}(a_i)
\Phi^{-1}_{ij}(E_k)
\phi^{(+)}(a_j)(H_0+|E_k|)^{-1}|\psi \rangle|| \\
\leq \lim_{k \rightarrow \infty}|E_k| \,
||{\sum_{i=1}^{N}\sum_{j=1}^{N}} |\Phi^{-1}_{ij}(E_k)| \;
||(H_0+|E_k|)^{-1} \phi^{(-)}(a_i)
\phi^{(+)}(a_j)(H_0+|E_k|)^{-1}|\psi \rangle|| \; ||\;,\eeqs
converges to zero. An explicit computation reveals that
\be \phi^{(+)}(a_j)(H_0+|E_k|)^{-1}|\psi \rangle =\sum_\sigma
{f_\sigma(a_j)\hat \psi(\sigma)\over
(\omega_\sigma+|E_k|)\omega_\sigma} |0 \rangle \;. \ee
The action of $(H_0+|E_k|)^{-1}\phi^{(-)}(a_i)$ onto this
expression leads to
\be (H_0+|E_k|)^{-1}\phi^{(-)}(a_i)\sum_\sigma {f_\sigma(a_j)\hat
\psi(\sigma)\over (\omega_\sigma+|E_k|)\omega_\sigma} |0 \rangle=
\sum_\sigma {f_\sigma(a_j)\hat \psi(\sigma) \over
(\omega_\sigma+|E_k|)\omega_\sigma}
\sum_{\sigma'}{f_{\sigma'}(a_i)\over
(\omega_{\sigma'}+|E_k|)}{a^\dagger_{\sigma'}\over
\sqrt\omega_{\sigma'}}|0 \rangle \;. \ee
We have now a term like $|\Phi^{-1}_{ij}(E_k)| \; ||
F(a_i,a_j|E_k)||$, which satisfies,
\be || \sum_{i,j=1}^{N} | \Phi^{-1}_{ij}(E_k)| \;  ||
F(a_i,a_j|E_k)|| \; || \leq \bigg[{\rm Tr}\, |\Phi^{-1}(E_k)|^2
\bigg]^{1/2} \bigg[{\rm Tr}\, ||F(E_k)||^2 \bigg]^{1/2} \;. \ee
This can be used in the above norm, and we get after one more use
of the Cauchy-Schwartz inequality,
\beqs & & \lim_{k \rightarrow \infty}
|E_k|||(H_0+|E_k|)^{-1}\phi^{(-)}(a_i) \Phi^{-1}_{ij}(E_k)
\phi^{(+)}(a_j)(H_0+|E_k|)^{-1}|\psi \rangle|| \cr &  & \leq N
|E_k| \max_{1\leq i,j \leq N } |\Phi^{-1}_{ij}(E_k)|
 \Bigg[\sum_{i=1}^{N}  \sum_{\sigma} {|f_\sigma(a_i)|^2 \over
(\omega_{\sigma}+|E_k|)^2 \omega_{\sigma}}\Bigg] \Big[\sum_\tau
{|\hat \psi(\tau)|^2\over \omega_\tau}\Big]^{1/2} \;, \eeqs
where we have used
\be \sum_\sigma {|f_\sigma(a_j)\hat \psi(\sigma)| \over
(\omega_\sigma+|E_k|)\omega_\sigma} \leq \Big[\sum_\sigma
{|f_\sigma(a_j)|^2\over
(\omega_\sigma+|E_k|)^2\omega_\sigma}\Big]^{1/2} \Big[\sum_\tau
{|\hat \psi(\tau)|^2\over \omega_\tau}\Big]^{1/2} \;. \ee
We recall that by choosing $k$ sufficiently large we can make the
off-diagonal elements as small as we like, while the diagonal
elements increase. Therefore, without repeating the arguments of
the previous section for sufficiently large values of $|E_k|$, we
can show that
\be \underset{1\leq i,j \leq N}\max \; |\Phi_{ij}^{-1}(E_k) | \leq
|| \Phi_{ij}^{-1}(E_k) || \leq  2 \underset{1\leq i \leq N}\max \;
|\Phi_{ii}^{-1}(E_k) | \ee
where
\be \max_{1\leq i \leq N} |\Phi_{ii}^{-1} (E_k)| \leq { C_{16}
\over \ln(|E_k|/(m-\mu_i^{min}))} \;. \ee
and the constant $C_{16}$ depends on the class of manifolds under
consideration. In the above equation $m$ should be superseded on
Cartan-Hadamard manifolds by its counterpart $m_{CH}$, as defined
in reference \cite{caglar}. Listed below are the values of this
constant for compact Riemannian manifolds with positive Ricci
curvature, for flat space and for Cartan-Hadamard manifolds.
\be C_{16}=\left\{
\begin{array}{lr} 2\pi & \text{for flat and compact manifolds} \\
\frac{2\pi}{c(\delta)} & \text{for Cartan-Hadamard manifolds}
\end{array} \right. \;. \ee
Thus, essentially we are faced with the sum/integral:
\be I = \sum_\sigma {|f_\sigma(a)|^2 \over
\omega_\sigma(\omega_\sigma+|E_k|)^2} \;. \ee
We will work this out: First we recall that
\be {1\over \omega_\sigma(\omega_\sigma+|E_k|)^2}={\Gamma(3)\over
\Gamma(2)\Gamma(1)} \int_0^1 \mathrm{d} \zeta \; {\zeta\over
[\omega_\sigma +\zeta |E_k|]^3} \;. \ee
Let us now use the exponential form for the integrand;
\be {\Gamma(3)\over \Gamma(2)\Gamma(1)} \int_0^1 \mathrm{d} \zeta
\; {\zeta\over [\omega_\sigma +\zeta |E_k|]^3} ={\Gamma(3)\over
2\Gamma(2)\Gamma(1)} \int_0^1 \mathrm{d} \zeta \; \zeta
 \int_0^\infty \mathrm{d} s \; s^2  e^{-s\omega_\sigma} e^{-\zeta s|E_k|} \;. \ee
and use subordination for $\omega_\sigma$,
\be e^{-s\omega_\sigma}= {s\over 2\sqrt{\pi}} \int_0^\infty
{\mathrm{d} u \over u^{3/2}} e^{-s^2/4u-m^2u} e^{-\lambda(\sigma)
u} \;, \ee
where $\lambda(\sigma)$ is the eigenvalue of the Laplacian defined
in \cite{caglar}. If we combine the last exponential with
$|f_\sigma(a)|^2$ terms, we get the heat kernel at the same
points, $K_u(a,a;g)$ and collecting them, we find
\beqs \nonumber I &=& {\Gamma(3)\over 4 \sqrt{\pi}
\Gamma(2)\Gamma(1)}\int_0^\infty {\mathrm{d} u \over u^{3/2}}
\int_0^1 \mathrm{d} \zeta \; \zeta \int_0^\infty \mathrm{d} s \;
s^3 \;  e^{-s^2/4u-m^2u}
K_u(a,a;g) e^{-\zeta s|E_k|} \\
\nonumber &= &  {\Gamma(3)\over 4\sqrt{\pi} \Gamma(2)\Gamma(1)}
\int_0^\infty {\mathrm{\mathrm{d}} u\over u^{3/2}}\int_0^\infty
\mathrm{d} s \; s^3 \; e^{-s^2/4u-m^2u} K_u(a,a;g)
\int_0^1 \mathrm{d} \zeta \; \zeta e^{-\zeta s|E_k|} \\
\nonumber &=& {\Gamma(3)\over 4\sqrt{\pi} \Gamma(2)\Gamma(1)}
\int_0^\infty {\mathrm{d} u\over u^{3/2}}\int_0^\infty \mathrm{d}
s \; s^3 \; e^{-s^2/4u-m^2u} K_u(a,a;g)
{1\over s^2 |E_k|^2} \Big[ 1-e^{-s|E_k|} -s|E_k| e^{-s|E_k|}\Big] \\
\nonumber &\leq & {\Gamma(3)\over 4\sqrt{\pi} |E_k|^2
\Gamma(2)\Gamma(1)} \int_0^\infty {\mathrm{d} u\over u^{3/2}}
\int_0^\infty \mathrm{d} s \; s \; e^{-s^2/4u-m^2u} \Big[{C_{17}
\over  A(\mathcal{M})}
+{C_{18} \over u }\Big] \Big[ 1-e^{-s|E_k|} -s|E_k|e^{-s|E_k|}\Big] \\
\nonumber&=& {1 \over |E_k|^2 } \int_0^\infty \mathrm{d} u \Big[
{C_{19} \over  A(\mathcal{M}) u^{3/2}} +{C_{20} \over
u^{5/2}}\Big] \int_0^\infty \mathrm{d} s \; s \;
e^{-s^2/4u-m^2 u} \Big[ 1-e^{-s|E_k|} -s|E_k|e^{-s|E_k|}\Big] \\
     &=& {1 \over |E_k|^2 } \int_0^\infty \mathrm{d} s \; s
\Big[ 1-e^{-s|E_k|} -s|E_k|e^{-s|E_k|}\Big]  \int_0^\infty
\mathrm{d} v \Big[{C_{19} m \over A(\mathcal{M}) v^{3/2}}+{C_{20}
m^3\over v^{5/2}}\Big] e^{-(m s)^2/4v-v} \;. \eeqs
where $A(\mathcal{M})$ is the area of the manifold. Here the most
divergent contribution comes from the last term in the above
expression, so we first analyze this term. By inspecting the
following integral representation of the modified Bessel function
$K_{3/2}(v)$ \cite{lebedev},
\be K_{3/2}(ms) = {1 \over 2} \Big( {m s \over 2} \Big)^{3/2}
\int_0^\infty {\mathrm{d} v \over v^{3/2+1}} e^{-(m s)^2/4v-v} \;,
\ee
we obtain the following
\be I_2 = {m^{3/2} C_{21} \over  |E_k|^2} \int_0^\infty
{\mathrm{d} s \over \sqrt{s}} K_{3/2}(m s) \Big[ 1-e^{-s|E_k|}
-s|E_k|e^{-s|E_k|}\Big] \;. \ee
We now use another integral representation of the modified Bessel
function $K_{3/2}(x)$ \cite{lebedev};
\be K_{3/2}(m s)= {\Gamma(2)2^{3/2} s^{3/2}\over \sqrt{\pi}
m^{3/2}} \int_0^\infty \mathrm{d} t  \; {\cos(m t)\over
(t^2+s^2)^2} \;. \ee
As a result we see that
\beqs \nonumber I_2 &=& {C_{22} \over |E_k|^2 } \int_0^\infty
\mathrm{d} s \; s  \int_0^\infty \mathrm{d} r \; s
{\cos(m s r)\over (r^2s^2+s^2)^2}\Big[ 1-e^{-s|E_k|} -s|E_k|e^{-s|E_k|}\Big] \\
\nonumber & \leq & {C_{22} \over |E_k|^2 } \int_0^\infty
{\mathrm{d} s \over s^2}  \Big[ 1-e^{-s|E_k|}
-s|E_k|e^{-s|E_k|}\Big]
\int_0^\infty {\mathrm{d} r \over (r^2+1)^2} \\
\nonumber & \leq & {C_{23} \over  |E_k|^2 } \int_0^\infty
{\mathrm{d} s \over s^{2}}
\Big[ 1-e^{-s|E_k|} -s|E_k|e^{-s|E_k|}\Big] \\
& \leq & {C_{23} \over  |E_k|} \int_0^\infty {\mathrm{d} s \over
s^{2}} \Big[ 1-e^{-s} -se^{-s}\Big] \;. \eeqs
We now note that the integral
\be \int_0^\infty {\mathrm{d} s \over s^{2}} \Big[ 1-e^{-s}
-se^{-s}\Big] \ee
is actually convergent. The first term instead becomes,
\beqs \nonumber I_1 &=& {2 \sqrt{\pi} C_{24} \over |E_k|^2 }
\int_0^\infty \mathrm{d} s \; \Big[ 1-e^{-s|E_k|} -s|E_k|e^{-s|E_k|}\Big] e^{-m s} \\
\nonumber & \leq & {C_{25} \over |E_k|^3 } \int_0^\infty
\mathrm{d} s \Big[ 1-e^{-s} -s e^{-s}\Big] e^{-m s /|E_k|}
 \\
& \leq & {C_{26} \over  A(\mathcal{M}) |E_k|^3 }\;. \eeqs
Hence,
\be \sum_\sigma {|f_\sigma(a)|^2 \over
\omega_\sigma(\omega_\sigma+|E_k|)^2} \leq {C_{27} \over
|E_k|}+{C_{26} \over A(\mathcal{M}) |E_k|^3}  \ee
is shown. As a result we see that
\be \lim_{k\to\infty}\,|| [1-|E_k| R(E_k)]|\psi \rangle || \leq
\lim_{k \to\infty} |E_k| \Big[{C_{28} \over |E_k|
\ln(|E_k|)}+{C_{29} \over A(\mathcal{M}) |E_k|^3
\ln(|E_k|)}\Big]\rightarrow 0 \;, \ee
which proves that our formula defines a densely defined closed
operator.
\section{Non-Relativistic Lee Model in Two and Three Dimensional Riemannian Manifolds}

\subsection{The Lower Bound on the Ground State Energy} \label{A Lower Bound on the Ground
State Energy for Two and Three Dimensions}

After the renormalization of the model in
\cite{nrleemodelonmanifold}, the principal operator was given
explicitly in three dimensions. We can similarly extend the
calculations given in three dimensions to the two dimensional case
\cite{ermanturgutlee2}, so that we have
\begin{eqnarray} \label{phi lee}
\Phi(E)&=&H_0-E+ \mu +\lambda^2 \int_0^\infty \mathrm{d} t \;
K_{t}(a,a;g)\left[e^{-t(m-\mu)} -e^{-t(H_0+m-E)}\right]\cr & \ &
\hskip-2cm - \; \lambda^2 \int_0^\infty \mathrm{d} t
\int_{\mathcal{M}^2} \mathrm{d}_{g}^{D} x \, \mathrm{d}_{g}^{D} y
\; K_{t}(x,a;g)K_{t}(y,a;g) \, \phi^\dag_g(x)
e^{-t(H_0+2m-E)}\phi_g(y) \;, \end{eqnarray}
where $D=2,3$ and $\mu$ is the experimentally measured bound state
energy of the system consisting of a boson and the attractive
fermion at the center. In this section we will restrict $E$ to the
real axis. In order to give the proof that the energy $E$ is
bounded from below, we split the principal operator as
\begin{equation}
\Phi(E) = K(E) - U(E) \;,
\end{equation}
such that
\begin{equation}
\label{K} K(E)= H_0 - E + \mu \;,
\end{equation}
and
\begin{eqnarray}
U(E) & = & U_1(E)+U_2(E)= - \lambda^2 \int_0^\infty \mathrm{d} t
\; K_{t}(a,a;g)\left[e^{-t(m-\mu)} -e^{-t(H_0+m-E)}\right]\cr & \
& \hskip-2cm + \; \lambda^2 \int_0^\infty \mathrm{d} t
\int_{\mathcal{M}^2} \mathrm{d}_{g}^{D} x \, \mathrm{d}_{g}^{D} y
\; K_{t}(x,a;g) \, K_{t}(y,a;g) \, \phi^\dag_g(x)\,
e^{-t(H_0+2m-E)} \, \phi_g(y) \;. \label{U}
\end{eqnarray}
It follows immediately that $K(E) \geq n m - E + \mu $, so it is a
positive definite operator from our assumption $E< n m + \mu$. Due
to the positivity of the heat kernel and since the difference of
the two exponentials is a positive operator, the first integral
term $U_1(E)$ is a negative operator. We thus remark that
\begin{equation}
U(E) \leq U_2(E) \;.
\end{equation}
This clearly forces
\begin{equation}
\Phi(E) \geq K(E)
 - U_2(E) \;,
\end{equation}
or rewriting it as
\begin{equation}
 \Phi(E) \geq K(E)^{1/2} \; \left(1-
\tilde{U_2}(E) \right) \; K(E)^{1/2}\;,
\end{equation}
where $\tilde{U}_2(E) = K(E)^{-1/2}\; U_2(E) \; K(E)^{-1/2}$ and
$K(E)$, $U_2(E)$ are positive operators (so is $\tilde{U}_2(E)$).
It must be emphasized that the unique square root of the positive
self-adjoint operators $K(E)$ are well defined for all real values
of $E$ below $\mu$. We will now show that by choosing $E$
sufficiently small it is always possible to make the operator
$\Phi(E)$ strictly positive, hence it becomes invertible, and has
no zeros beyond this particular value of $E$ (in the last section,
the self-adjointness will be further clarified). Therefore, if we
impose
\begin{equation}
||\tilde{U}_{2}(E)|| < 1 \;, \label{strictpositivity}
\end{equation}
then the principal operator $\Phi(E)$ becomes strictly positive.
For Cartan-Hadamard manifolds, we have obtained in
\cite{nrleemodelonmanifold}
\beqs ||\tilde{U}_{2}(E)|| \leq n \; C_{30} m^{D/2} \; { \lambda^2
\, \Gamma(2)\over \Gamma(1/2)^2} (nm + \mu -E)^{{D \over 2}-2}
\Gamma(2-{D \over 2}) \Bigg[ {\sqrt{\pi} \Gamma(1-{D \over 4})
\over \Gamma({3\over 2}- {D \over 4})}\Bigg]^2 \;. \label{upper
bound U2 Cartan hadamard} \eeqs
Then the strict positivity of the principal operator
(\ref{strictpositivity}) implies a lower bound for the ground
state energy
\begin{equation}
E_{gr} \geq n m  + \mu  - \Bigg( n C_{31} \lambda^2
m^{D/2}\Bigg)^{1 \over 2-{D \over 2}} \label{energycartan}\;,
\end{equation}
where %
\begin{equation}
C_{31} = C_{30} \; { \pi \Gamma(2) \Gamma(1-{D \over 4})^2
\Gamma(2-{D\over 2}) \over \Gamma({1 \over 2})^2 \Gamma({3\over
2}-{D \over 4})^2} \;.
\end{equation}
For the compact manifolds with Ricci curvature bounded from below
by $-K \geq 0$, we have similarly obtained
\begin{eqnarray}
||\tilde{U}_{2}(E)||& \leq & n \; {\lambda^2 \, \Gamma(2)\over
\Gamma(1/2)^2} \bigg[{4 \over V(\mathcal{M}) \mu^{D \over 2}} + {4
A'^{1/2} m^{D/4} \pi^{1/2} \Gamma(2-{D\over 4}) \Gamma(1-{D\over
4}) \over \mu^{D \over 4} V(\mathcal{M})^{1/2} \Gamma({3 \over 2}-
{D \over 4 })} \cr & & + \;  {A' m^{D/2} \pi \Gamma(2-{D \over 2})
\Gamma(1-{D \over 4})^2 \over \Gamma({3 \over 2}-{D \over 4})^2}
 \bigg] {1 \over (n m + \mu - E)^{2-{D \over 2}}} \;,
\label{upper bound U compact} \end{eqnarray}
so the lower bound of the ground state energy was found
\begin{equation}
E_{gr} \geq n m + \mu - \left( n \lambda^2 C_{32} \right)^{1 \over
2-{D \over 2}} \label{energyh4}\;,
\end{equation}
where
\beqs & & C_{32} = {\Gamma(2) \over \Gamma(1/2)^2} \bigg[{4 \over
V(\mathcal{M}) \mu^{D \over 2}} + {4 A'^{1/2} m^{D/4} \pi^{1/2}
\Gamma(2-{D\over 4}) \Gamma(1-{D\over 4}) \over \mu^{D \over 4}
V(\mathcal{M})^{1/2} \Gamma({3 \over 2}- {D \over 4 })} \cr & &
\hspace{5cm} + \; {A' m^{D/2} \pi \Gamma(2-{D \over 2})
\Gamma(1-{D \over 4})^2 \over \Gamma({3 \over 2}-{D \over 4})^2}
\bigg] \;. \eeqs
Therefore, the lower bounds on the ground state energies for
different classes of manifolds (\ref{energycartan}) and
(\ref{energyh4}) are of almost the same form up to a constant
factor, so the form of the lower bound has a general character.

\subsection{Existence of the Hamiltonian for the Lee Model in Two
and Three Dimensional Riemannian Manifolds}

The explicit formula for the resolvent of the Hamiltonian in terms
of the inverse of the principal operator $\Phi^{-1}(E)$ is given
in \cite{nrleemodelonmanifold, ermanturgutlee2} by
\be
R(E)= {1 \over H -E}= \left(%
\begin{array}{cc}
  \alpha & \gamma \\
  \beta & \delta \\
\end{array}%
\right)\;, \ee
where
\beqs \alpha &=&{1\over H_0-E}+{1\over H_0-E} \; b^{\dag} \;
\Phi^{-1} (E) \; b \; {1\over H_0-E}\cr \beta &=&- \Phi^{-1}(E)\;
b \; {1\over H_0-E}\cr \gamma & = & - {1\over H_0-E} \; b^{\dag}
\; \Phi^{-1}(E) \cr
 \delta &=&\Phi^{-1}(E)\cr b &=&\lambda \phi_g(a)
\;. \eeqs
Let us check that the resolvent identity
$R(E_1)-R(E_2)=(E_1-E_2)R(E_1)R(E_2)$ is satisfied, that is, we
must have
\beqs & &
\left(%
\begin{array}{cc}
  \alpha(E_1)-\alpha(E_2) & \gamma(E_1)-\gamma(E_2) \\
  \beta(E_1)-\beta(E_2) & \delta(E_1)-\delta(E_2) \\
\end{array}%
\right) \cr & & = (E_1-E_2) \left(%
\begin{array}{cc}
  \alpha(E_1)\alpha(E_2) +  \gamma(E_1) \beta(E_2) & \alpha(E_1)\gamma(E_2) + \gamma(E_1)\delta(E_2) \\
  \beta(E_1) \alpha(E_2) + \delta(E_1) \beta(E_2) & \beta(E_1) \gamma(E_2) + \delta(E_1)\delta(E_2) \\
\end{array}%
\right) \;. \label{resolvent equation lee} \eeqs
We first consider the first diagonal element of the above matrix.
Using the fact that free resolvent satisfies the resolvent
identity, we get
\be \begin{split} R_0(E_1) b^{\dag} \Phi^{-1}(E_1) \Bigg[
\Phi(E_1)-\Phi(E_2) & + b (R_0(E_1)-R_0(E_2)) b^{\dag} \\ & +
E_1-E_2 \Bigg] \Phi^{-1}(E_2) b R_0(E_2)  = 0 \end{split} \;. \ee
Let us look at the term in the square bracket more closely. By
using the explicit expression of the principal operator (\ref{phi
lee}), this term becomes
\beqs & & \lambda^2 \int_0^\infty \mathrm{d} t \;
K_{t}(a,a;g)\left[e^{-t(H_0+ m-E_2)} -e^{-t(H_0+m-E_1)}\right] \cr
& &  + \; \lambda^2 \int_0^\infty \mathrm{d} t
\int_{\mathcal{M}^2} \mathrm{d}_{g}^{D} x \, \mathrm{d}_{g}^{D} y
\; K_{t}(x,a;g)K_{t}(y,a;g) \, \phi^\dag_g(x) \Bigg[e^{-t(H_0
+2m-E_2)}- e^{-t(H_0 +2m-E_1)} \Bigg] \phi_g(y) \cr & & + \;
\lambda^2 \phi_g(a) \Bigg[(H_0-E_1)^{-1}-(H_0-E_2)^{-1} \Bigg]
\phi^{\dag}_{g}(a) \label{remaining} \;. \eeqs
One can shift the operator $\phi^{\dag}_{g}(x)$ to the left
\be {1\over H_0-E} \phi^{\dag}_{g}(x)= \int_{\mathcal{M}}
\mathrm{d}_{g}^{D} x' \; \phi^{\dag}_{g}(x') \int_0^\infty
\mathrm{d} t \; e^{-t (H_0+m-E)} \, K_{t}(x,x';g) \;,
\label{normalorderingcreation} \ee
and shift the operator $\phi_{g}(x)$ to the right
\be \phi_{g}(x) {1\over H_0-E} = \int_{\mathcal{M}}
\mathrm{d}_{g}^{D} x' \; \int_0^\infty \mathrm{d} t \;
e^{-t(H_0+m-E)} \, K_{t}(x,x';g) \phi_{g}(x') \;,
\label{normalorderingannihilation}  \ee
which we have also used in \cite{nrleemodelonmanifold} for the
renormalization. The last term in the equation (\ref{remaining})
can be normal ordered as
\be
\begin{split}
& \lambda^2 \int_0^\infty \mathrm{d} t \;
K_{t}(a,a;g)\left[e^{-t(H_0+ m-E_1)} -e^{-t(H_0+m-E_2)}\right] \\
& + \lambda^2 \int_0^\infty \mathrm{d} t \int_{\mathcal{M}^2}
\mathrm{d}_{g}^{D} x \, \mathrm{d}_{g}^{D} y \;
K_{t}(x,a;g)K_{t}(y,a;g) \, \phi^\dag_g(x) \Bigg[e^{-t(H_0
+2m-E_1)}- e^{-t(H_0 +2m-E_2)} \Bigg] \phi_g(y) \label{remaining2}
\end{split} \;. \ee
Then we prove that
\be \Phi(E_1)-\Phi(E_2) + b (R_0(E_1)-R_0(E_2)) b^{\dag} + E_1-E_2
= 0 \;. \label{identity} \ee
The other term in the matrix equality (\ref{resolvent equation
lee})
\be \gamma(E_1) - \gamma(E_2) =(E_1-E_2)
\Bigg[\alpha(E_1)\gamma(E_2) - \gamma(E_1) \delta(E_2) \Bigg]\ee
can be written as
\be - R_0(E_1) b^{\dag} \Phi^{-1}(E_1) \Bigg[\Phi(E_1)-\Phi(E_2) +
b (R_0(E_1)-R_0(E_2)) b^{\dag} + E_1-E_2 \Bigg] \Phi^{-1}(E_2)=0
\;, \ee
due to (\ref{identity}). Similarly, the other terms can be put
into the following forms
\beqs & & \Phi^{-1}(E_1) \Bigg[\Phi(E_1)-\Phi(E_2) + b
(R_0(E_1)-R_0(E_2)) b^{\dag} + E_1-E_2 \Bigg] \Phi^{-1}(E_2) b
R_0(E_2) = 0 \cr & & \Phi^{-1}(E_1) \Bigg[\Phi(E_1)-\Phi(E_2) + b
(R_0(E_1)-R_0(E_2)) b^{\dag} + E_1-E_2 \Bigg]\Phi^{-1}(E_2)= 0 \;,
\eeqs
and they are all satisfied thanks to the equality
(\ref{identity}). Hence, we prove that the resolvent identity is
satisfied.

Recall that the resolvent for the Lee model is defined in the
following Fock space $\mathcal{F_B}^{(n+1)}(\mathcal{H}) \otimes
\chi_+ \oplus \mathcal{F_B}^{(n)}(\mathcal{H}) \otimes \chi_-  $,
for any given $n \in \mathbb{N}$, and $\chi_{\pm}$ is the spin
states. In matrix form, we have $R(E):
\mathcal{F_B}^{(n+1)}(\mathcal{H}) \oplus
\mathcal{F_B}^{(n)}(\mathcal{H})  \rightarrow
\mathcal{F_B}^{(n+1)}(\mathcal{H}) \oplus
\mathcal{F_B}^{(n)}(\mathcal{H}) $. Then we must show that
\be ||E_k R(E_k) |f \rangle + |f \rangle || = |||E_k| R(E_k) |f
\rangle - |f \rangle || \rightarrow 0 \;, \ee
as $\textcolor[rgb]{0.00,0.00,0.50}{k} \rightarrow \infty$. Here
$|f \rangle \in \mathcal{F_B}^{(n+1)}(\mathcal{H}) \oplus
\mathcal{F_B}^{(n)}(\mathcal{H}) $ and the norm is taken with
respect to $\mathcal{F_B}^{(n+1)}(\mathcal{H}) \oplus
\mathcal{F_B}^{(n)}(\mathcal{H})$. Let us decompose the vector
$|f\rangle$ as
\be
\left(%
\begin{array}{c}
  |f^{(n+1)} \rangle \\
   |f^{(n)} \rangle  \\
\end{array}%
\right) \ee
where
\be |f^{(n)} \rangle = \int_{\mathcal{M}^n} \mathrm{d}_g^D x_1
\ldots \mathrm{d}_g^D x_n \; f(x_1,x_2, \ldots, x_n)
|x_1,x_2,\ldots,x_n \rangle \;. \ee
So we have
\be
\begin{split} &
\Bigg|\Bigg| \left(%
\begin{array}{cc}
  |E_k| \alpha(E_k) & |E_k| \gamma(E_k) \\
  |E_k| \beta(E_k) & |E_k| \delta(E_k) \\
\end{array}%
\right) \left(%
\begin{array}{c}
  |f^{(n+1)} \rangle \\
  |f^{(n)} \rangle  \\
\end{array}%
\right) - \left(%
\begin{array}{c}
  |f^{(n+1)} \rangle \\
  |f^{(n)} \rangle  \\
\end{array}%
\right) \Bigg|\Bigg| \\ & = \Bigg[|||E_k| \alpha(E_k) |f^{(n+1)}
\rangle - |f^{(n+1)} \rangle + |E_k| \gamma(E_k) |f^{(n)} \rangle
||^2 \\ & + || |E_k| \beta(E_k)|f^{(n+1)} \rangle + |E_k|
\delta(E_k) |f^{(n)} \rangle - |f^{(n)} \rangle ||^2 \Bigg]^{1/2}
  \\ & \leq \Bigg[ \Bigg( |||E_k| \alpha(E_k)
|f^{(n+1)} \rangle - |f^{(n+1)} \rangle || + || |E_k| \gamma(E_k)
|f^{(n)} \rangle ||\Bigg)^2 \\ & + \Bigg( || |E_k|
\beta(E_k)|f^{(n+1)} \rangle || + |||E_k| \delta(E_k) |f^{(n)}
\rangle - |f^{(n)} \rangle || \Bigg)^2 \Bigg]^{1/2} \;, \label{all
terms existence} \end{split} \ee
since $||A+B|| \leq ||A|| + ||B||$. We shall investigate each norm
separately. Let us first consider the term $|| |E_k|
\beta(E_k)|f^{(n+1)} \rangle ||$
\be
\begin{split}
|| \lambda |E_k| \Phi^{-1}(E_k) \phi_g(a) {1 \over H_0 + |E_k|}
|f^{(n+1)}\rangle|| \leq \lambda |E_k| ||\Phi^{-1}(E_k)  || ||
\phi_g(a) {1 \over H_0 + |E_k|} |f^{(n+1)}\rangle ||
\end{split} \;.
\ee
Using the formula (\ref{normalorderingannihilation}) for
$E=-|E_k|$ and $x=a$, we get
\be
\begin{split} ||
\phi_g(a) {1 \over H_0 + |E_k|} |f^{(n+1)}\rangle || & \leq ||
\int_{\mathcal{M}} \mathrm{d}^{D}_{g} x \; \int_{0}^{\infty}
\mathrm{d} t \; e^{-t|E_k|} K_t(x,a;g)\phi_g(x) |f^{(n+1)} \rangle
|| \\ & \leq \Bigg[\int_{\mathcal{M}} \mathrm{d}^{D}_{g} x \;
\Bigg(\int_{0}^{\infty} \mathrm{d} t \; e^{-t|E_k|} K_t(x,a;g)
\Bigg)^2 \Bigg]^{1/2} \sqrt{n+1} ||\; |f^{(n+1)} \rangle || \;.
\end{split}
\ee
Let us first consider the compact manifolds. Then, one can take
the integral over the variable $t$ by the help of the upper bound
of the heat kernel (\ref{heatkernel bound}) for compact manifolds
and obtain
\be
\begin{split} || \phi_g(a) & {1 \over H_0 + |E_k|}
|f^{(n+1)}\rangle||  \leq \sqrt{n+1} || |f^{(n+1)} \rangle ||\\ &
\hskip-3cm \times \Bigg[\int_{\mathcal{M}} \mathrm{d}^{D}_{g} x \;
  \bigg({m d^2(x,a) \over |E_k|}\bigg)  \Bigg( {C_{33} \over V(\mathcal{M})} K_1 \left(2
d(x,a)\sqrt{m |E_k|/C_3}\right)
\\ & \hspace{1cm} + C_{34} \left({ m |E_k|\over d^2(x,a)} \right)^{D/4} K_{{D \over
2}-1}\left(2 d(x,a)\sqrt{m |E_k|/C_3}\right)\Bigg)^2 \Bigg]^{1/2}
\;.
\end{split}
\ee
We now choose Riemann normal coordinates around the point $a$,
assuming that $\delta < \mathrm{inj}(a)$. Then, we split the
integration region into the two parts as $\int_{\mathcal{M}} =
\int_{B_{\delta}(a)} + \int_{\mathcal{M}\setminus B_{\delta}(a)}$.
Expressing the first integral in the Gaussian spherical
coordinates, we get
\be
\begin{split} || \phi_g(a) & {1 \over H_0 + |E_k|}
|f^{(n+1)}\rangle|| \leq \sqrt{n+1} || |f^{(n+1)} \rangle || \\
& \times \Bigg[\int_{\mathbb{S}^{D-1}} \mathrm{d} \Omega
\int_{0}^{\delta} \mathrm{d} r \;
 \;  r^{D+1}
\; \Bigg( { m A_{+}^{D-1}(K_1,0) \over |E_k|} \Bigg) \Bigg[
{C_{33} \over V(\mathcal{M})} K_1\left(2 r \sqrt{m
|E_k|/C_3}\right)
\\ & + C_{34} \left({ m |E_k|\over r^2} \right)^{D/4} K_{{D \over
2}-1}\left(2 r \sqrt{m |E_k|/C_3}\right) \Bigg]^2 \\ & +
\int_{\mathcal{M}\setminus B_\delta(a)} \mathrm{d}^{D}_{g} x \;
\Bigg( { m d^2(x,a) \over |E_k|} \Bigg)  \Bigg[ {C_{33} \over
V(\mathcal{M})} K_1\left(2 d(x,a)\sqrt{m |E_k|/C_3}\right)
\\ & +  C_{34} \left({ m |E_k|\over d^2(x,a)} \right)^{D/4} K_{{D \over
2}-1}\left(2 d(x,a)\sqrt{m |E_k|/C_3}\right)\Bigg]^2 \Bigg]^{1/2}
\;,
\end{split} \label{lee existence 1}
\ee
where we have used the equations (\ref{J}) and (\ref{Aplus
Aminus}). Let us now consider the first integral. It is smaller
than the following expression
\be
\begin{split} &
\int_{\mathbb{S}^{D-1}} \mathrm{d} \Omega \int_{0}^{\infty}
\mathrm{d} r \;
 \; r^{D+1}
\; \Bigg( {m A_{+}^{D-1}(K_1,0)  \over |E_k|} \Bigg) \Bigg[{C_{33}
\over V(\mathcal{M})} K_1\left(2 r \sqrt{m |E_k|/C_3}\right)
\\ & +
C_{34} \left({ m |E_k|\over r^2} \right)^{D/4} K_{{D \over
2}-1}\left(2 r \sqrt{m |E_k|/C_3}\right) \Bigg]^2 \;.
\end{split}
\ee
One can evaluate the integrals \cite{gradshteyn}
\be
\begin{split}
\int_{0}^{\infty} \mathrm{d} r \;
 \; r^{D+1} K_{1}^{2}(a r) & = {\sqrt{\pi} \Gamma(1 + {D \over 2}) \Gamma(2 + {D \over 2}) \Gamma({D \over 2})
 \over 4 a^{D+2} \Gamma({3+D \over 2})} \\ \int_{0}^{\infty} \mathrm{d} r \;
 \; r \; K_{{D \over 2}-1}^{2}(a r) & =  {\pi (D-2) \csc(\pi D/2) \over 4 a^{2}}
 \\ \int_{0}^{\infty} \mathrm{d} r \;
 \; r^{{D \over 2}+1} \; K_{1}(a r) K_{{D \over 2}-1}(a r) & = {2^{D \over 2}
  \Gamma({D \over 2}) \over (D+2) a^{{D \over 2}+2} } \;,
\end{split} \label{besselestimates}
\ee
where $a \in \mathbb{R}^+$ and $D=2,3$. Then the upper bound of
the first integral in (\ref{lee existence 1}) becomes
\be  {m A_{+}^{D-1}(K_1,0) \over |E_k|} \Bigg( {C_{35} \over
V^2(\mathcal{M})}(m|E_k|)^{-{(D+2) \over 2}} + C_{36} (m|E_k|)^{{D
\over 2}-1} + {C_{37} \over V(\mathcal{M})} (m|E_k|)^{-1} \Bigg)
\label{first term} \;. \ee
For Cartan-Hadamard manifolds, we do not repeat the analysis above
because the upper bound of the heat kernel for Cartan-Hadamard
manifolds given in the equation (\ref{heatkernel bound})
corresponds to removing the volume term from the one for the
compact manifolds. As a result, we get the upper bound of the
first term in the equation (\ref{lee existence 1}) for
Cartan-Hadamard manifolds
\be
\begin{split} {m C_{38} \over |E_k|} (m|E_k|)^{{D \over 2}-1}
\end{split} \;. \label{1st term CH}
\ee
Let us now consider the second term in the equation (\ref{lee
existence 1}) for compact and Cartan-Hadamard manifolds. Due to
the upper bounds of the Bessel functions used in
\cite{teomanfatih}, we find for compact manifolds
\be
\begin{split} &
\int_{\mathcal{M}\setminus B_\delta(a)} \mathrm{d}^{D}_{g} x \;
\Bigg( { m d^2(x,a) \over |E_k|} \Bigg)  \Bigg[ {C_{33} \over
V(\mathcal{M})} K_1\left(2 d(x,a)\sqrt{m |E_k|/C_3}\right)
\\ & \hspace{5cm} +  C_{34} \left({ m |E_k|\over d^2(x,a)} \right)^{D/4} K_{{D \over
2}-1}\left(2 d(x,a)\sqrt{m |E_k|/C_3}\right)\Bigg]^2 \\ & \leq
\int_{\mathcal{M}\setminus B_\delta(a)} \mathrm{d}^{D}_{g} x \;
\bigg({m d^2(x,a)\over |E_k|}\bigg) \Bigg[ {C_{33} \over
V(\mathcal{M})} \exp \bigg( {- d(x,a) \sqrt{m|E_k|/C_3}}\bigg)
\\ & \times \Bigg( {1 \over 2 d(x,a) \sqrt{m |E_k|/C_3}} + {1 \over 2} \Bigg)
+ C_{34} \left({ m |E_k|\over d^2(x,a)} \right)^{D/4}  {2 \exp
\bigg( - {2 d(x,a) \over (4-D)} \sqrt{m|E_k|/C_3} \bigg) \over (2
d(x,a)\sqrt{m|E_k|/C_3})^{(4-D)/2}}  \Bigg]^2 \;.
\end{split} \label{lee existence 2}
\ee
Since $d(x,a) \geq \delta $ for all $x \in \mathcal{M}\setminus
B_\delta(a)$, the upper bound of the above equation is
\be
\begin{split} & \exp \bigg( - \delta \sqrt{m|E_k|/C_3}  \bigg)
\int_{\mathcal{M}\setminus B_\delta(a)} \mathrm{d}^{D}_{g} x \;
\bigg({m d^2(x,a) \over |E_k|} \bigg) \Bigg[ {C_{33} \over
V(\mathcal{M})}  \exp \bigg( - {d(x,a)\over 2} \sqrt{m |E_k|/C_3}
\bigg) \\ & \times \Bigg( {1 \over 2 \delta \sqrt{m|E_k|/C_3}} +
{1 \over 2} \Bigg)
 +   C_{34} \left({ m |E_k|\over \delta^2} \right)^{D/4} {2 \over (2 \delta \sqrt{m|E_k|/C_3})^{(4-D)/2}} \\ &
 \hspace{6cm} \times \exp \Bigg( \left({d(x,a) \over 2}- {2 d(x,a)
\over (4-D)} \right) \sqrt{m|E_k|/C_3} \Bigg) \Bigg]^2 \;.
\end{split} \label{lee existence 3} \ee
For compact manifolds, we have a simplification. This upper bound
above is smaller than
\be
\begin{split}
& \exp \bigg(- \delta \sqrt{m |E_k|/C_3} \bigg) \int_{\mathcal{M}}
\mathrm{d}^{D}_{g} x \; \bigg({m d^2(x,a) \over |E_k|}\bigg)\Bigg[
{C_{33} \over V(\mathcal{M})} \Bigg( {1 \over 2 \delta \sqrt{m
|E_k|/C_3}} + {1 \over 2} \Bigg)
\\ & \hspace{3cm} +   C_{34} \left({ m |E_k|\over \delta^2} \right)^{D/4}
{2 \over (2 \delta \sqrt{m|E_k|/C_3})^{(4-D)/2}} \Bigg]^2 \;.
\end{split} \label{lee existence 4}
\ee
Due to the fact the geodesic distance between any two points on
the manifold and the volume of the manifold is finite, that is,
$d(x,a)\leq d_{max}(a)=\max_{x}d(x,a)$, the upper bound to the
above integral can easily be found as
\be
\begin{split}
& \bigg({m d^{2}_{max}(a) \over |E_k|} \bigg) \exp  \bigg( {-
\delta \sqrt{m|E_k|/C_3}} \bigg) V(\mathcal{M}) \Bigg[ {C_{33}
\over V(\mathcal{M})}
\Bigg( {1 \over 2 \delta \sqrt{m |E_k|/C_3}} + {1 \over 2} \Bigg) \\
& \hspace{7cm} + C_{34} \left({ m |E_k|\over \delta^2}
\right)^{D/4} {2 \over (2 \delta \sqrt{m |E_k|/C_3})^{(4-D)/2}}
\Bigg]^2 \;.
\end{split}  \label{lee existence 5}
\ee
For Cartan-Hadamard manifolds, we similarly find
\be
\begin{split} & \leq C_{39} m^{D-1} |E_k|^{D-3} \exp  \bigg( {- \delta \left(m |E_k|/ C_5 \right)^{1/2}}
\bigg) \int_{\mathcal{M}\setminus B_\delta(a)} \mathrm{d}^{D}_{g}
x \;  { \exp \bigg[ {2({d(x,a) \over 2}- {2 d(x,a) \over (4-D)})
\sqrt{m |E_k| /C_5}} \bigg] \over d^{2}(x,a)} \\ & \leq {C_{39}
m^{D-1} |E_k|^{D-3} \exp  \bigg( {- \delta \sqrt{m |E_k|/C_5}}
\bigg) \over \delta^2} \int_{\mathcal{M}} \mathrm{d}^{D}_{g} x \;
\exp \bigg[ {-d(x,a) \left({4 \over 4-D}-1 \right) \sqrt{m
|E_k|/C_5}} \bigg] \;,
\end{split}  \label{2nd term CH} \ee
where we have used $d(x,a) \geq \delta$ for all $x \in
\mathcal{M}\setminus B_\delta(a)$. Let us write the above integral
in Gaussian spherical coordinates as we did in Section 2,
\be
\begin{split} \int_{\mathbb{S}^{D-1}} \mathrm{d} \Omega \int_{0}^{\rho_{\Omega}}
\mathrm{d} r \; r^{D-1} J(r,\theta) \exp \bigg[ {-r \left({4 \over
4-D}-1 \right) \sqrt{m|E_k|/C_5}} \bigg] \;.
\end{split} \ee
To proceed further we assume that $\mathcal{M}$ has \textit{Ricci
tensor bounded from below by $K_1$}. As a result of this and using
the equations (\ref{J}) and (\ref{sn}), the upper bound to the
equation (\ref{2nd term CH}) becomes
\be
\begin{split} & \leq {C_{40}
m^{D-1} |E_k|^{D-3} \exp  \bigg( {- \delta \sqrt{m |E_k|/C_5}}
\bigg) \over \delta^2 (-K_1)^{(D-1)/2}}  \int_{0}^{\infty}
\mathrm{d} r \; \sinh^{D-1}(\sqrt{-K_1}r) \\ & \hspace{7cm} \times
\exp \bigg[ {-r \left({4 \over 4-D}-1 \right) \sqrt{m |E_k|/C_5}}
\bigg] \;.
\end{split}  \ee
Since $\sinh^{D-1}(x) \leq e^{(D-1)x}/2^{D-1}$, we can take the
integral and get
\be {C_{41} m^{D-1}  |E_k|^{D-3} e^{- \delta \sqrt{m |E_k|/C_5}}
\over (-K_1)^{(D-1)/2} \delta^2 \Bigg[\left({4 \over 4-D}-1
\right) \sqrt{ m |E_k|/C_5} - (D-1) \sqrt{-K_1} \Bigg]} \;,
\label{lee existence 6} \ee
as long as $\Bigg[\left({4 \over 4-D}-1 \right) \sqrt{m|E_k|/C_5}
- (D-1) \sqrt{-K_1}\Bigg] \geq 0$. But this is always satisfied
for sufficiently large values of $|E_k|$.

Therefore, if we combine the results (\ref{first term}) and
(\ref{lee existence 5}) we obtain for compact manifolds that
\be
\begin{split} &
|| \phi_g(a) {1 \over H_0 + |E_k|} |f^{(n+1)}\rangle || \leq
\sqrt{n+1} || |f^{(n+1)} \rangle || \Bigg[ {m A_{+}^{D-1}(K_1,0)
\over |E_k|} \Bigg( {C_{33} \over
V^2(\mathcal{M})}(m|E_k|)^{-{(D+2) \over 2}} \\ & + C_{34}
(m|E_k|)^{{D \over 2}-1} + {C_{35} \over V(\mathcal{M})}
(m|E_k|)^{-1} \Bigg) + \bigg({m d^{2}_{max}(a) \over |E_k|} \bigg)
\exp  \bigg( {- \delta \sqrt{m|E_k|/C_3}} \bigg)
\\ & \times V(\mathcal{M}) \Bigg( {C_{33} \over V(\mathcal{M})}
\Bigg( {1 \over 2 \delta \sqrt{m |E_k|/C_3}} + {1 \over 2} \Bigg)
+ C_{34} \left({ m |E_k|\over \delta^2} \right)^{D/4} {2 \over (2
\delta \sqrt{m |E_k|/C_3})^{(4-D)/2}} \Bigg)^2 \Bigg]^{1/2} \;,
\end{split}  \ee
and the results (\ref{1st term CH}) and (\ref{lee existence 6})
for Cartan-Hadamard manifolds give
\be
\begin{split} &
|| \phi_g(a) {1 \over H_0 + |E_k|} |f^{(n+1)}\rangle || \leq
\sqrt{n+1} || |f^{(n+1)} \rangle || \Bigg[
 {m C_{38} \over |E_k|} (m|E_k|)^{{D \over 2}-1} \\ & + {C_{41} m^{D-1}  |E_k|^{D-3}
e^{- \delta \sqrt{m |E_k|/C_5}} \over (-K_1)^{(D-1)/2} \delta^2
\bigg(\left({4 \over 4-D}-1 \right) \sqrt{ m |E_k|/C_5} - (D-1)
\sqrt{-K_1} \bigg)} \Bigg]^{1/2} \;.
 \end{split}  \ee
We are now going to find an upper bound of the inverse norm of the
principal operator. In order to do this, let us recall that we
split the principal operator when we try find the lower bound of
the ground state energy. We now split the principal operator in
the following way: $\Phi=(K-U_1)-U_2$, where $U_1$ and $U_2$ are
defined exactly as before. Then, we have
\beqs &  \Phi^{-1} = \left(K-U_1\right)^{-1/2} \; \left[1-
\left(K-U_1\right)^{-1/2} U_2 \left(K-U_1\right)^{-1/2}
\right]^{-1} \; \left(K-U_1\right)^{-1/2} \;. \eeqs
Let us substitute the identity operator $K^{1/2}K^{-1/2}$ between
the operators $\left(K-U_1\right)^{-1/2}$ and $U_2$. Hence,
\beqs & \Phi^{-1} = \left(K-U_1\right)^{-1/2}\left[1- X
\right]^{-1} \; \left(K-U_1\right)^{-1/2} \;, \eeqs
where we have defined $X= \left(K-U_1\right)^{-1/2}K^{1/2}
\tilde{U}_2 K^{1/2}\left(K-U_1\right)^{-1/2}$ for simplicity. Here
the following operator can be written as an infinite geometric sum
\be \left[1- X \right]^{-1} =\sum_{l=0}^{\infty} X^l \;, \ee
as long as $||X||<1$. This leads to
\beqs ||\left[1- X \right]^{-1}|| \leq \left[1- ||X|| \right]^{-1}
\;. \eeqs
Since $-U_1$ is a positive operator, $(K-U_1)^{-1/2} \leq
K^{-1/2}$. Then, we have
\beqs ||X|| \leq  ||\tilde{U}_2|| \;. \eeqs
If we make $|E_k|$ sufficiently large then $||\tilde{U}_2|| \leq
1/2$ and
\be \left[1- ||X|| \right]^{-1} \leq 2 \;. \ee
As a result of this, we get
\be ||\Phi^{-1}(E_k)|| \leq  2 ||K^{-1/2}(E_k)||^2 \leq {2 \over
|E_k|} \;, \ee
where we have used
\be K^{-1/2}(E_k)= \left( H_0 + \mu + |E_k| \right)^{-1/2} \leq {1
\over |E_k|^{1/2}} \;. \ee
Then, substituting the equations (\ref{first term}) and (\ref{lee
existence 5}) for compact or substituting the equations (\ref{1st
term CH}) and (\ref{lee existence 6}) for Cartan-Hadamard
manifolds into the equation (\ref{lee existence 1}) and using the
above upper bound for the inverse principal operator, and taking
the limit as $k \rightarrow \infty$, we eventually obtain
\be
\begin{split} |E_k| \; ||\Phi^{-1}(E_k)|| \; || \phi_g(a) {1 \over H_0 + |E_k|}
|f^{(n+1)}\rangle|| \rightarrow 0
\end{split} \;.
\ee
Let us consider the other terms in the equation (\ref{all terms
existence}) now:
\be
\begin{split}
& |||E_k| \alpha(E_k) |f^{(n+1)} \rangle - |f^{(n+1)} \rangle ||
\leq || \left({|E_k| \over H_0 + |E_k|}-1 \right)|f^{(n+1)}
\rangle || \\ & + \lambda^2 |E_k| \; || {1 \over H_0 + |E_k|}
\phi^{\dag}_{g}(a)|| \; || \Phi^{-1}(E_k) || \; || \phi_g(a) {1
\over H_0 + |E_k|} | f^{(n+1)} \rangle ||
\end{split} \;.  \ee
The \textit{upper} bound for the norm $|| {1 \over H_0 + |E_k|}
\phi^{\dag}_{g}(a)||$ can be similarly found, and comes out to  be
the same as the one for $|| \phi_{g}(a) {1 \over H_0 + |E_k|}
||$(the norms are different in general). As a result of this, we
obtain
\be
\begin{split}
& |||E_k| \alpha(E_k) |f^{(n+1)} \rangle - |f^{(n+1)} \rangle ||
\rightarrow 0
\end{split} \;, \ee
as $k \rightarrow \infty$. Similarly, the following term
\be \begin{split} |||E_k| \gamma(E_k)|f^{(n)} \rangle || \leq
\lambda |E_k| \; ||  {1 \over H_0 + |E_k|}\phi_{g}^{\dag}(a)|
f^{(n)} \rangle || \; || \Phi^{-1}(E_k) ||
\end{split} \ee
and
\be \begin{split} |||E_k| \beta(E_k)|f^{(n+1)} \rangle || \leq
\lambda |E_k| \; || \phi_{g}(a) {1 \over H_0 + |E_k|} | f^{(n+1)}
\rangle || \; || \Phi^{-1}(E_k) ||
\end{split} \ee
both vanishes as $k \rightarrow \infty$. Moreover, we have
\beqs & || \left( |E_k| \delta(E_k) -1 \right) |f^{(n)} \rangle ||
=  || \left[ |E_k| \; \Phi^{-1}(E_k) -1 \right] |f^{(n)} \rangle
|| \cr & =  || \left[ |E_k| \;K^{-1/2}(E_k)\left[ 1+
(1-\tilde{U}(E_k))^{-1}-1 \right] K^{-1/2}(E_k) -1 \right]
|f^{(n)} \rangle || \cr & \hskip-0.7cm \leq || \left(
|E_k|K^{-1}(E_k)-1 \right) |f^{(n)} \rangle || + |||E_k|
K^{-1/2}(E_k) \left[(1-\tilde{U}(E_k))^{-1}-1 \right]
K^{-1/2}(E_k)|f^{(n)} \rangle || \cr & \hskip-0.7cm = || \left(
|E_k|K^{-1}(E_k)-1 \right) |f^{(n)} \rangle || + |||E_k|
K^{-1/2}(E_k) (1-\tilde{U}(E_k))^{-1} \tilde{U}(E_k)
K^{-1/2}(E_k)|f^{(n)} \rangle || \;, \label{123} \eeqs
where we have used the fact that the factor
$(1-\tilde{U}(E_k))^{-1}$ can be considered as an infinite
geometric sum. The first term goes to zero as $k \rightarrow
\infty$ since
\be
\begin{split}
& || \left( {|E_k| \over H_0 + |E_k| + \mu }-1 \right) |f^{(n)}
\rangle || = || \left( {|E_k|+\mu-\mu \over H_0  + |E_k|+ \mu}-1
\right) |f^{(n)} \rangle || \\ & \leq || \left( {|E_k|+\mu \over
H_0 + |E_k|+ \mu }-1 \right) |f^{(n)} \rangle || + || \left({\mu
\over H_0  + |E_k|+ \mu} \right) |f^{(n)} \rangle ||
\\ & \leq || \left( {|E_k|+\mu \over H_0 + |E_k|+
\mu}-1 \right) |f^{(n)} \rangle || + {\mu \over |E_k|} || |f^{(n)}
\rangle || \;,
\end{split}
\ee
where the term containing $H_0$ vanishes as $k \rightarrow \infty$
because it is the free resolvent and the second part clearly goes
to zero. This shows that the first term in the equation
(\ref{123}) vanishes in the limit. As for the second term, it is
smaller than
\beqs |E_k| ||K^{-1/2}(E_k) || \;
\left[(1-||\tilde{U}(E_k)||)^{-1} \right] \;
\left[||\tilde{U}_1(E_k)||+ ||\tilde{U}_2(E_k)|| \right] ||
K^{-1/2}(E_k)||\; || |f^{(n)} \rangle || \;. \eeqs
Since $m > \mu$ for bound states, one can easily see that
\be  \begin{split}&  ||\tilde{U}_1(E_k)|| \leq C_{42} \lambda^2
||(H_0+m+ |E_k|)^{-1/2} \; \int_0^\infty \mathrm{d} t \;
K_{t}(a,a;g)\left[e^{-t(m-\mu)} -e^{-t(H_0+m+ |E_k|)}\right]
 \\ & \hspace{9cm} \times  (H_0+m+|E_k|)^{-1/2}|| \\ & \leq
\begin{cases} C_{43} \lambda^2
||(H_0+m+ |E_k|)^{-1} \ln \left( {H_0  + m + |E_k| \over m-\mu}
\right) || & \mathrm{for} \; D=2
\\ C_{44} \lambda^2
||(H_0+m+|E_k|)^{-1/2}|| & \mathrm{for} \; D=3  \end{cases} \;.
 \end{split}  \ee
Here we use the fact that the operator in the parenthesis, which
we call $A(s)$  is positive, and for a positive family, if two
integrable functions satisfy $0 \leq f(s) \leq g(s)$, then $\int
\mathrm{d} s \; f(s) A(s) \leq \int \mathrm{d} s \; g(s) A(s)$.
Moreover, for positive operators, order relation implies the same
ordering for the norms of the operators. For simplicity, we have
also disregarded the more convergent in $|E_k|$ coming from the
volume terms of upper bound of the heat kernel for compact
manifolds. We now note that
\be \begin{split}
    ||{1\over H_0+|E_k|+m}\ln\left({ H_0+|E_k|+m \over m-\mu}\right)
    ||& =||\int_0^1 {\mathrm{d} t \over [H_0+|E_k|+m]t+m-\mu}||
    \\
& \leq ||\int_0^1 {\mathrm{d} t \over [H_0+|E_k|+m]t^2+m-\mu}||\\
& \leq ||{1 \over \sqrt{ m-\mu}
[H_0+|E_k|+m]^{1/2}}||\int_0^\infty {\mathrm{d} s \over
s^{2}+1} \\
 & \leq  {C_{45} \over \sqrt{ m-\mu} |E_k|^{1/2}}
\;. \end{split} \ee
Hence, we get
\be ||\tilde{U}_1(E_k)|| \leq  \begin{cases} {C_{46} \lambda^2
\over \sqrt{ m-\mu} |E_k|^{1/2}}
 & \mathrm{for} \; D=2
\\ {C_{47} \lambda^2 \over |E_k|^{1/2}}
 & \mathrm{for} \; D=3  \end{cases} \;.
 \label{U1} \ee
Using the results (\ref{upper bound U2 Cartan hadamard}) and
(\ref{upper bound U compact}) for $E=-|E_k|$ with the above
analysis, we finally obtain
\be
\begin{split} &
|E_k| || K^{-1}(E_k)|| \; || \tilde{U}(E_k)||
 \; || (1-\tilde{U}(E_k))^{-1} || \; || |f^{(n)} \rangle ||
 \rightarrow 0
\end{split} \;,
\ee
as $k \rightarrow \infty$ and this completes the proof that our
renormalized formula corresponds to the resolvent of a densely
defined closed operator.

We will further show  that $\Phi(E)$ is a holomorphic self-adjoint
family of type (A) in the sense of Kato \cite{kato}. This will in
turn justify the claim that the resolvent corresponds to a
self-adjoint operator. To prove this we will use the theorem given
by R. W{\"u}st \cite{wust}. First, we define a holomorphic family
of type (A) as follows: Let $G\subset \mathbb{C}$ be domain and
$L(z)$ be a family of {\it closed} linear operators, acting on a
Hilbert space ${\cal H}$,  $\{ L(z) | z\in G\}$. If
\beqs & 1) & \mathrm{the \; domain} \;
\mathcal{D}(L(z))=\mathcal{D} \; \mathrm{is \; independent \;of}
\;  z\in G \cr & 2) & \mathrm{for \; any} \; f \in \mathcal{D}, \;
\mathrm{and} \; g \in \mathcal{H} \; \mathrm{then} \; \langle g
|L(z)| f \rangle \; \mathrm{is \; holomorphic \; in} \; G \;,
\eeqs
then this is a {\it holomorphic family of type (A)}.

An operator which is a holomorphic family of type (A) is a
{\it self-adjoint} holomorphic family of type (A) if
\beqs & 1) & G \; \mathrm{is \;  a \; {\it symmetric \; domain} \;
of \; the \; complex \; plane \; relative \; to \; the \; real \;
axis.} \cr & 2) & \mathcal{D} \; \mathrm{is \; {\it dense} \; in}
\; {\cal H} \cr & 3) &
\mathcal{D}(L(z)^\dagger)=\mathcal{D}(L(z^*)) \; \mathrm{for \;
all} \; z\in G \;. \eeqs
{\bf Theorem}(W\"ust): Let $G$ be a symmetric domain of complex
plane relative to the real axis, and $L(z)$
is a holomorphic family of type (A)  defined on $G$.  Assume
\beqs & 1) & \mathcal{D} \; \mathrm{is \; dense \; in} \; {\cal H}
\cr & 2) & \mathcal{D}(L(z)^\dagger) \supset \mathcal{D}(L(z^*))
\;. \eeqs
Let
\be M=\{ z\in G| L(z)^\dagger=L(z^*)\} \;. \ee
If $M$ {\it is not the empty set then it is the whole domain }
$G$. This implies that the family is a self-adjoint holomorphic
family of type (A).

Let us consider our case. We choose the domain $G$ as
\be G= \{ E \in \mathbb{C} | \Re(E) < \mu \} \;, \ee
which is symmetric with respect to the real axis. The principal
operator $\Phi(E)$ given explicitly in (\ref{phi lee}) formally
satisfies the relation $\Phi(E)^\dagger=\Phi(E^*)$ so this implies
$\mathcal{D}(\Phi(E)^\dagger)\supset \mathcal{D}(\Phi(E^*))$. Let
us assume that the family is holomorphic for now. Note that a
densely defined holomorphic family of operators satisfying the
formal relation $\Phi(E)^\dagger=\Phi(E^*)$ is closable. Proof:
Let us consider the common domain $\mathcal{D}$, and choose $|g_l
\rangle\in \mathcal{D} \rightarrow 0$ as $l \rightarrow \infty$.
We further assume that $\Phi(E)|g_l\rangle$ converges to some
$|g(E)\rangle$. Then we have
\be \langle f| \Phi(E) g_l\rangle=\langle
\Phi^\dag(E)f|g_l\rangle=\langle \Phi(E^*) f|g_l\rangle \to 0 \;,
\ee
for any $|f \rangle\in \mathcal{D}$ as $l\rightarrow \infty$. This
implies that
\be \Phi(E)|g_l\rangle \rightarrow|g(E)\rangle=0 \;. \ee
This is the requirement of closure. Of course,  we must  establish
this closure uniformly, that is,  we need to show that for every
sequence $|g_l\rangle $ converging to some element, if
$\Phi(E_0)|g_l\rangle $ converges for one $E_0$ inside the region
$G$, then it converges for all $E \in G$. Hence, we can define a
unique closure over $G$, the closures having a common domain
$\mathcal{D}^{-}$. Once we determine a common domain for the
family $\Phi(E)$, we will prove that there is indeed a closure
over a common domain.

For the family $\Phi(E)$, we choose
$\mathcal{D}=\mathcal{D}(H_0)$. It is well known that if ${\cal
M}$ is a geodesically complete manifold then Laplacian defined on
$\mathcal{M}$ is a closed, densely defined self-adjoint operator
\cite{gaffney1, gaffney2}. Then, the operator $H_0=
\int_{\mathcal{M}} \mathrm{d}_g^{D}x \; \phi^{\dagger}_{g}(x)(-{1
\over 2m} \nabla_g^2) \phi_g(x)$ defined over the direct products
of $n$  copies of the Hilbert space $L^2(\mathcal{M})$ is also a
densely defined (essentially) self-adjoint operator. Moreover, the
finite direct sum of such operators will preserve this property.

The first term $H_0-E+\mu$ is obviously defined over this domain
$\mathcal{D}(H_0)$ and it is a closed operator. Note that the
other term can be defined by the spectral theorem,
\be \int_0^\infty  \mathrm{d} s \;
K_s(a,a;g)[e^{-s(m-\mu)}-e^{-s(H_0+ m -E)}] \ee
and it is a positive operator when $E$ is {\it real} and $\Re (E)
<\mu$.  Its domain of definition  includes $\mathcal{D}(H_0)$ when
$\Re (E) <\mu$. To see this we will use a different integral
representation,
\beqs & & \int_0^\infty \mathrm{d} s \; K_s(a,a;g) [e^{-s(m-\mu)}-
e^{-s(H_0-E+m)}] \cr & & \hspace{2cm} = \int_0^\infty \mathrm{d} s
\; s \; K_s(a,a;g)  \int_0^1 \mathrm{d} u \;
e^{-su(H_0+\mu-E)-s(m-\mu)}(H_0-E+\mu) \;. \eeqs
When the operator acts on an element $|f^{(n)}\rangle$ in the
domain of $H_0$, the norm of the resulting vector is smaller than,
\be \int_0^\infty \mathrm{d} s \; s \; K_s(a,a;g) \int_0^1
\mathrm{d} u \; ||e^{-su(H_0+\mu-E)-s(m-\mu)}|| ||(H_0-E+\mu)
|f^{(n)}\rangle|| \;. \ee
Now we can estimate the following factor using the bounds on the
heat kernels given in (\ref{heatkernel bound}),
\beqs & & \int_0^\infty \mathrm{d} s \; s K_s(a,a;g) \int_0^1
\mathrm{d} u  ||e^{-su(H_0+\mu-E)-s(m-\mu)}|| \cr & & \leq
\int_0^1 \mathrm{d} u \int_0^\infty \mathrm{d} s \; s
\left[{C_1\over V({\cal M})}+{C_2\over (s/2m)^{D/2}}\right] e^{-s
u n m } e^{-su(\mu-\Re(E))-s(m-\mu)} \;. \eeqs
Thus we show that
\be ||\int_0^\infty \mathrm{d} s \; K_s(a,a;g) [
e^{-s(H_0-E+m)}-e^{-s(m-\mu)}]|| \leq F(\mu-\Re(E)) \Big( ||
H_0|f^{(n)}\rangle||+|\mu-E| || |f^{(n)}\rangle||\Big) \;,\ee
where
\beqs & & F(\mu-\Re(E)) = {C_1 \over (nm + \mu -\Re(E))
V(\mathcal{M})} \left( {1 \over m- \mu} +  {1 \over (n+1)m-
\Re(E)} \right) \cr & & + {{C_2}(2m)^{D/2} \Gamma(2-{D \over 2})
\over (nm + \mu -\Re(E)({D \over 2}-1)} \left[ ((n+1)m-
\Re(E))^{{D \over 2}-1} -(m-\mu)^{{D \over 2}-1} \right] \;. \eeqs
As a result the domain of this operator family includes
$\mathcal{D}(H_0)$. In fact,  by the spectral theorem the
operators so defined are closed, when we restrict them to a
smaller domain, i. e. to $\mathcal{D}(H_0)$ they remain closed.
 So the sum of the two pieces, $H_0+\mu-E$ and the term above,
 defined over $\mathcal{D}(H_0)$ is closed, since they were
 already closed operators defined over a common domain.

The last part requires more work,  for  this we will first show
that $U(E)$ is relatively bounded with respect to $H_0$ hence its
domain includes  $\mathcal{D}(H_0)$. Moreover, if we have a
holomorphic family of operators defined over a dense domain, then
they are preclosed, that is we can define the closure of this
family, as we have shown. It is easy to see that
\be ||U(E)H_0^{-1} H_0 |f^{(n)} \rangle || \leq ||U(E)H_0^{-1}||\
|| H_0 |f^{(n)} \rangle|| \;,\ee
where the first norm can be estimated by exactly the same method
developed in \cite{nrleemodelonmanifold}. So we are giving the
result in order not to repeat the similar calculations, for $n\geq
1$,
\beqs
 ||U(E)H_0^{-1}|| &\leq& C_{48} n \int_0^\infty \mathrm{d} s\int_0^1 \mathrm{d} u  \;
 s\;
 K^{1/2}_{us}(a,a;g)K_s^{1/2}(a,a;g) e^{-snm} e^{su \Re(E)}
\;. \eeqs
After using the upper bound of the heat kernel given in
(\ref{heatkernel bound}) and defining new variables $p=
C_1/V(\mathcal{M})$ and $q(s)= C_2/(s/2m)^{D/2}$ we get
\beqs & &
 ||U(E)H_0^{-1}|| \leq C_{48} n \int_0^\infty \mathrm{d} s\int_0^1 \mathrm{d} u  \;
 s \;
 e^{-snm} e^{su \Re(E)} \Bigg[ p^2 + p q(s) + {p q(s) \over u^{3/2}} + {q(s)^2 \over u^{3/2}}
 \cr & &  \hspace{5cm} + {2 p q(s) \over u^{3/4}} - {2 p q(s) \over
 u^{3/4}}\Bigg]^{1/2} \cr & & \hskip-0.5cm \leq C_{48} n \int_0^\infty \mathrm{d} s\int_0^1 \mathrm{d} u  \;
 s \;
 e^{-snm} e^{su \Re(E)} \Bigg[ \Bigg(p + {q(s) \over u^{3/4}}\Bigg) +  \sqrt{pq(s)}
 \Bigg(1 + {1 \over u^{3/2}} - {2 \over
 u^{3/4}}\Bigg) \Bigg]
\;. \eeqs
Taking the $s$ and $u$ integral, we obtain
\beqs  ||U(E)H_0^{-1}|| \leq {n C_{49} \over
V(\mathcal{M})(nm-\mu)^2}+ {n (2m)^{D/2} C_{50} \over
(nm-\mu)^{2-D/2}} + {n (2m)^{D/4} C_{51} \over
\sqrt{V(\mathcal{M})} (nm-\mu)^{2-D/4}} \;, \eeqs
since $\Re{E}<\mu$. Thus we choose the domain of $U(E)$ as
$\mathcal{D}(H_0)$, and now the family is closable  over this
domain. However, as a result of the closure, the domains for
different values of $E$ may become different. In fact, this does
not happen, as we will see.

Now we show that we can perform the closure {\it uniformly}, as a
result of the following: for any $E_1, E_2\in G$
$\Phi(E_1)-\Phi(E_2)$ becomes a bounded operator. A short
computation shows that,
\be || \Phi(E_1)-\Phi(E_2)|| \leq |E_1-E_2|\Big[1+(n+1)\lambda^2
\int_0^\infty \mathrm{d} s \; s K_s(a,a;g) e^{-s n m}
e^{-s(m-\mu)}\Big] \;. \ee
If $|g_l\rangle \in \mathcal{D}$ is convergent to a vector
$|f\rangle $, and assume that $\Phi(E_1)|g_l\rangle $ converges to
$|g(E_1)\rangle $  for one $E_1$, then we set $\Phi(E_1)|f
\rangle= |g(E_1) \rangle$ to define the closure at point $E_1$.
Then, for any $E_2$, we have
\beqs
 &\ &  ||\Phi(E_2)|g_l\rangle-\Phi(E_2)|f\rangle || =
  ||\Phi(E_2)|g_l\rangle -|g(E_1)\rangle -[\Phi(E_2)-\Phi(E_1)]|f\rangle||\cr
 &\ &\ \ =||[\Phi(E_2)-\Phi(E_1)]|g_l>+\Phi(E_1)|g_l\rangle -|g(E_1)\rangle-[\Phi(E_2)-\Phi(E_1)]|f\rangle||\cr
 &\ &\ \ <
||[\Phi(E_2)-\Phi(E_1)]||\, |||g_l
\rangle-|f\rangle||+||\Phi(E_1)|g_l\rangle-|g(E_1)\rangle||\mapsto
0\;, \eeqs
and this shows that whenever $|g_l \rangle$ converges to
$|f\rangle$ and $\Phi(E_1)|g_l \rangle$ converges to $|g(E_1)
\rangle$, we have $\Phi(E_2)|g_l\rangle$ becomes convergent and
the resulting vector is exactly equal to $\Phi(E_2)|f\rangle$ as
it should be for the requirements of the closure. Hence the sum of
all these three parts will make a holomorphic family $\Phi(E)$
with a dense common domain $\mathcal{D}(H_0)$. Moreover, the sum
is {\it closable over a dense common domain} which  we call
$\mathcal{D}(H_0)^-$.

We would now make holomorphicity more precise, up to now we have
not actually made use of it. To prove that the family is
holomorphic we will refer to the following theorem, which is
stated in a slightly simplified form according  to our needs and
the proof of which can be found in \cite{everitt}: Assume $X$ is a
measure space with a $\sigma$-finite measure $\nu$ defined on it,
let $I$ be a measurable subset of $X$. Let $G$ be a open domain of
the complex plane. Consider a function $\gamma: I \times G\mapsto \mathbb{C} $ such
that
\beqs
&1)& \ \gamma(x,.)\in L^1(X, |\nu|)\cr
&2)&\ \gamma(.,z)\ \ \ {\rm is\ holomorphic\  in}\ G\cr
& 3)& \
\int_I |\mathrm{d}\nu| |\gamma(x,z)| \ \ {\rm is\ bounded\ on \
all \ compact \ subsets \ of } \ G \;. \eeqs
Then the function
\begin{equation}
\Gamma(z)=\int_I \mathrm{d}\nu \;\gamma(x,z) \ \ {\rm is \
holomorphic \ in\ } \ G \;.
\end{equation}
To use the above theorem, let us write our family in the following
form $\Phi(E)=V(E)(H_0-E+\mu)=[1+V_1(E)+V_2(E)](H_0-E+\mu)$, where
\beqs V_1(E)&=&\lambda^2 (H_0-E+\mu)^{-1} \int_0^\infty \mathrm{d}
s \; K_s(a,a;g)[e^{-s(m-\mu)}-e^{-s(H_0-E+m)}]\cr V_2(E)&=&
\lambda^2 \int_0^\infty \mathrm{d} s \; \phi^{\dag}_{g}(x)
K_s(x,a;g) e^{-s(H_0-E+2m)} K_s(a,y;g) \phi_{g}(y)
(H_0-E+\mu)^{-1} \;. \eeqs
By using  our previous estimate in (166) and (167) we see that,
the first term $V_1(E)$ is indeed uniformly bounded in
$\Re{E}<\mu$. Hence, the integrals $\langle f^{(n)}|V_1(E)|g^{(n)}
\rangle$ for all functions $f,g$ in the Fock space are absolutely
convergent. Moreover, any such matrix element satisfies all the
other conditions on
 holomorphicity and integrability.
 The second part,  again using ideas very similar to the previous estimates,  can be written as
\beqs V_2(E)=\lambda^2 \int_0^\infty \mathrm{d} s \; s \int_0^1
\mathrm{d} u \;
\phi^{\dagger}_{g}(K_{su}(.,a;g))e^{-s(H_0-E+m+(1-u)\mu + m u)}
\phi_g(K_s(.,a;g)) \;.\eeqs
This integrand as a function of $E$ is holomorphic in $E$ for
$\Re{E}<\mu$ and it is  absolutely integrable for any $\Re{E}<\mu$
on $[0,\infty)\times [0,1]$. Similarly, it can be shown that the
following bound holds
\beqs |\langle f^{(n)}|V_2(E)|g^{(n)} \rangle| \leq n {C_{52}
\over (nm-\Re{E})^{D/2-1}} |||f^{(n)})\rangle|| |||g^{(n)}\rangle|| \;, \eeqs
which clearly shows that for $\Re{E}< \mu$ is uniformly bounded
everywhere, hence on compact subsets as well. Hence, applying the
theorem stated above we see that the resulting function $V_2(E)$
is holomorphic for $\Re{E}<\mu$. There is one subtle point about
the closure operation, but this is also solved by the following
observation. Let us consider the  limit of $\Phi(E)|g_l\rangle$ as
$l \to \infty$ as a function of $E$ for any convergent $|g_l
\rangle$ sequence in the closure operation. If this sequence is
uniformly convergent on compact subsets the limit is a holomorphic
function (by an application of Morera's theorem). Note that this
family $\Phi(E)|g_l\rangle$ is norm bounded by a constant multiple
of a simple function given in the equation (171), $|E_1-E_2|$.
This function itself is uniformly bounded on compact sets centered
around any given point $E_1$, hence the sequence of functions
$\Phi(E)|g_l\rangle$ is a uniformly convergent sequence. This
shows that the closure remains a holomorphic function of $E$ for
$\Re{E}<\mu$ as required. Thus we complete the proof that the
closure of the family $\Phi(E)$ over $D(\mathcal{H}_0)^{-}$ is
holomorphic for $\Re(E)<\mu$.

Now we are ready to apply the theorem of W\"ust. If we choose
$E\in \mathbb{R}$ and sufficiently small $E<E_*$, then $U(E)$ has
relative bound with respect to $K(E)$ which is less than 1. Hence,
by Kato-Rellich theorem \cite{kato} of perturbations of
self-adjoint operators, $U(E)$ will be self- adjoint for $E<E_*$.
By the theorem of W\"{u}st, the family is self-adjoint everywhere
as desired. This  result is important to establish that the
spectrum only lies along the real axis,
 and justifies our search for the lower bound of energy and
 shows that the resulting operator is self-adjoint as it should be.

\section{Conclusion}
\label{conclusion}

In this paper, we have proven that for the three models that we
have constructed, namely non-relativistic point interactions in
two and three dimensional Riemannian manifolds, relativistic point
interactions in two dimensional Riemannian manifolds and
non-relativistic Lee model in two and three dimensional Riemannian
manifolds, the Hamiltonian after renormalization is a densely
defined self-adjoint operator.

\section{Acknowledgments}

\c{C}. Dogan and O. T. Turgut would like to thank to Prof. M.
Znojil and Prof. P. Exner for the kind invitation to the Doppler
Institute in Prague and for various discussions related to the
subject under consideration. O. T. Turgut has two times visited
the Department of Mathematics of KTH, Stockholm during the
completion of this work and would like to thank Prof. J. Hoppe for
his kind invitations and his continuous support.

\end{document}